\documentclass[prc,showpacs,showkeys,superscriptaddress,twocolumn]{revtex4-1}

\usepackage{amsmath,bm}
\usepackage{graphicx}
\usepackage{epstopdf}
\usepackage{subfigure}
\usepackage{epsfig}
\usepackage{amsmath,amssymb,amsfonts}
\usepackage{enumitem}
\usepackage{xcolor}
\usepackage[utf8]{inputenc}
\usepackage{verbatim}
\usepackage{array}
\usepackage{hyperref}
\graphicspath{{figures/}} 
\definecolor{theblue}{RGB}{0,50,230}
\hypersetup{
  colorlinks=true,
  linkcolor=theblue,
  citecolor=theblue,
  urlcolor=theblue
}

\begin{document}

\renewcommand{\topfraction}{0.85}
\renewcommand{\textfraction}{0.1}
\renewcommand{\floatpagefraction}{0.75}
\def\be{\begin{eqnarray}}
\def\ee{\end{eqnarray}}
\newcommand{\mt}[1]{\textrm{\tiny #1}}
\def\pt{{p_\mt{T}}}
\def\ponet{{p_\mt{T1}}}
\def\ptwot{{p_\mt{T2}}}

\title{Substructure grooming of inclusive and photon-tagged jets in heavy-ion collisions}

\author{Sa Wang}
\email{wangsa@ctgu.edu.cn}
\affiliation{College of Science, China Three Gorges University, Yichang 443002, China}
\affiliation{Center for Astronomy and Space Sciences and Institute of Modern Physics, China Three Gorges University, Yichang 443002, China}
\affiliation{Key Laboratory of Quark \& Lepton Physics (MOE) and Institute of Particle Physics, Central China Normal University, Wuhan 430079, China}

\author{Shuang Li}
\email{lish@ctgu.edu.cn}
\affiliation{College of Science, China Three Gorges University, Yichang 443002, China}
\affiliation{Center for Astronomy and Space Sciences and Institute of Modern Physics, China Three Gorges University, Yichang 443002, China}

\author{Jin-Wen Kang}
\affiliation{Key Laboratory of Quark \& Lepton Physics (MOE) and Institute of Particle Physics, Central China Normal University, Wuhan 430079, China}

\author{Ben-Wei Zhang}
\email{bwzhang@mail.ccnu.edu.cn}
\affiliation{Key Laboratory of Quark \& Lepton Physics (MOE) and Institute of Particle Physics, Central China Normal University, Wuhan 430079, China}

\author{Enke Wang}
\affiliation{State Key Laboratory of Nuclear Physics and Technology, Institute of Quantum Matter, South China Normal University, Guangzhou 510006, China}
\affiliation{Guangdong Basic Research Center of Excellence for Structure and Fundamental Interactions of Matter, Guangdong Provincial Key Laboratory of Nuclear Science, Guangzhou 510006, China}

\date{\today}

\begin{abstract}
Jet substructure provides a powerful probe of partonic interactions within the quark-gluon plasma (QGP) in heavy-ion collisions. In this paper, we present a systematic theoretical study of the groomed substructures for both inclusive jets and photon-tagged jets ($\gamma+$jets) utilizing the Dynamical and Soft-Drop Grooming algorithms in PbPb collisions by employing the SHELL transport model. Our theoretical calculations exhibit a suppression at high $k_{\rm T,g}$, the relative transverse momentum between the two subjets in the groomed substructure, consistent with the recent ALICE measurements. We show that the suppression of high $k_{\rm T,g}$ arises from the combined effects of the reduction of the subleading subjet transverse momentum due to partonic energy loss and the narrowing of the groomed jet radius $R_g$ induced by selection bias. Our findings demonstrate that no enhancement is observed at high $k_{\rm T,g}$, even in the complete absence of selection bias. Furthermore, we propose that the broadening of $R_g$ in photon-tagged jets, which are less susceptible to selection bias compared to inclusive jets, provides relatively direct evidence of the jet substructure broadening. Our analysis reveals that the $R_g$ broadening becomes more pronounced as the jet radius increases, where the medium-induced gluon radiation plays a dominant role in driving such broadening. In particular, we find that as the jet radius increases, the Soft Drop grooming algorithm exhibits a better resolving power for the contribution of the medium response to the jet substructure broadening.
\end{abstract}

\maketitle

\section{Introduction}
\label{sec:introduction}

High-energy nuclear collisions provide a unique opportunity to study the deconfined nuclear matter experimentally, the quark-gluon plasma (QGP), at the Relativistic Heavy Ion Collider (RHIC) and the Large Hadron Collider (LHC). Exploring the properties of QGP is of crucial importance for understanding the behavior of quantum chromodynamics (QCD) under extreme conditions~\cite{Matsui:1986dk, Wang:1991hta, Gyulassy:2004zy, PHENIX:2004vcz, STAR:2005gfr, Chen:2024aom, Shou:2024uga, Wu:2025xpz, Zhu:2025xiz, Ma:2026xiz}. Jet quenching phenomena, the strong interaction between high-$p_{\rm T}$ jets and the QGP medium, stand out as one of the most effective probes for unraveling the properties of this new state of matter~\cite{Braaten:1991we, Baier:1996kr, Baier:1996sk, Zakharov:1996fv, Gyulassy:1999zd, Wiedemann:2000za, Arnold:2001ba, Arnold:2001ms, Wang:2001ifa, Vitev:2009rd, Gyulassy:1993hr, Gyulassy:2003mc, Qin:2015srf, Xie:2024xbn, Yang:2025yaf, Kong:2024wdk, Kang:2023qxb, He:2022tod, Barata:2022krd, Andres:2022ndd}.

As a complementary and promising approach to the study of full-jet observables, jet substructures possess the potential to reveal the detailed mechanisms of jet-medium interactions at finer resolution scales, such as the medium-induced gluon radiation \cite{Baier:1996kr, Baier:1996sk}, medium response \cite{KunnawalkamElayavalli:2017hxo, Pablos:2019ngg, Chen:2020tbl, Casalderrey-Solana:2020rsj, He:2018xjv, Ke:2020clc}, and the ``Moli\`{e}re scattering'' \cite{DEramo:2012uzl, DEramo:2018eoy}. A comprehensive overview of the advancements on this topic can be found in references~\cite{Marzani:2019hun, Cunqueiro:2021wls, Apolinario:2022vzg, Sorensen:2023zkk, Arslandok:2023utm, Mehtar-Tani:2025rty}. It is noted that jet formation is influenced by various factors, including initial-state radiation, hadronization, multiple parton interactions, and the underlying event~\cite{Larkoski:2014wba}. Non-perturbative effects involved in these processes theoretically limit the ability of pQCD to describe the substructure of jets. Jet grooming techniques have been developed to remove the influence of soft, wide-angle radiation, thereby reducing the contamination of non-perturbative contributions and preserving the characteristics of hard splittings during the jet formation~\cite{Krohn:2009th, Ellis:2009su, Butterworth:2008iy}.

Jet grooming techniques also build a bridge between pQCD calculations and experimental measurements in heavy-ion collisions~\cite{Chien:2016led, Mehtar-Tani:2016aco, Milhano:2017nzm, Caucal:2019uvr, Chang:2017gkt, Li:2017wwc, Ringer:2019rfk, Caucal:2021cfb, Cunqueiro:2022svx, Wang:2022yrp, JETSCAPE:2023hqn, Tachibana:2025rcx}. Recently, the ALICE collaboration has reported the measurements on the nuclear modification of jet $k_{\rm T,g}$, the relative transverse momentum of the two subjets in the groomed jet substructure~\cite{ALICE:2024fip}. However, the measured $k_{\rm T,g}$ distribution does not exhibit the naively expected enhancement due to jet-medium interactions at the high-$k_{\rm T,g}$ region; instead, a suppression is observed. It is conjectured that this suppression arises from the ``selection bias''~\cite{Baier:2001yt, Renk:2012ve, Wang:2021jgm, Brewer:2021hmh}, the same factor that causes the observed narrowing of the jet angle difference between jet axes $\Delta R_{\rm axis}$~\cite{ALICE:2023dwg} and the jet girth~\cite{ALICE:2018dxf, Yan:2020zrz}. The mechanism by which selection bias influences the nuclear modification of $k_{\rm T,g}$ still lacks a thorough understanding. On the other hand, compared to inclusive jets, the vector boson tagged jets ($Z^0/\gamma+$jets)~\cite{CMS:2024zjn, Wang:2024plm} and hadron+jets~\cite{STAR:2023pal, ALICE:2023qve} have been demonstrated to significantly mitigate the impact of selection bias which obscures the direct observation of the real nuclear modification effects on jet substructure. Investigating such tagged jets using grooming techniques would allow for a more direct examination of how jet substructure is modified in nucleus-nucleus collisions when selection bias is substantially reduced.

This paper presents a theoretical study of the groomed substructures for both inclusive jets and $\gamma+$jets in heavy-ion collisions. Firstly, we carry out the nuclear modification of $k_{\rm T,g}$ distributions for inclusive jets in PbPb collisions compared to the recent ALICE data. We demonstrate that the suppression of high $k_{\rm T,g}$ results from the combined effect of the reduction of subleading subjet transverse momentum ($p_{\rm T,2}$) due to partonic energy loss and the narrowing of jet radius ($R_g$) induced by selection bias. We show that even in the complete absence of selection bias, no enhancement can be observed at high $k_{\rm T,g}$. Furthermore, we propose that the broadening of $R_g$ in photon-tagged jets will provide relatively direct evidence of jet-medium interactions. We systematically investigate the impact of different grooming algorithms, parameter settings, jet cone size, and jet-medium interaction mechanisms on the nuclear modification of $R_g$ for $\gamma+$jets. Finally, we compute the nuclear modification of $R_g$ for $\gamma+$jets at both the parton and hadron levels, to test the sensitivity of different grooming algorithms to hadronization effects.

The paper is structured as follows. In Sec.~\ref{sec:ppbaseline}, we discuss the groomed jet substructure of inclusive jets in pp collisions. In Sec.~\ref{sec:framework}, we introduce the theoretical framework for simulating the jet evolution in nucleus-nucleus collisions. Furthermore, we show and discuss the main results of this paper in Sec.~\ref{sec:res}. Finally, we conclude this paper in Sec.~\ref{sec:sum}.

\section{Dynamical and Soft-Drop Grooming of Jet Substructure in pp collisions}
\label{sec:ppbaseline}

QCD jets produced in hadron collisions exhibit a rich internal structure, which potentially reflects the characteristics of the parton shower following the hard scattering. However, jet substructure is also influenced by various non-perturbative effects, which limit the predictive power of pQCD. For this reason, grooming techniques have been developed to remove soft or wide-angle radiation, thereby reducing the impact of non-perturbative processes. The earliest jet grooming algorithms include trimming~\cite{Krohn:2009th}, pruning~\cite{Ellis:2009su}, and the mass drop tagger (MDT)~\cite{Butterworth:2008iy}. The Soft-Drop Grooming (SDG)~\cite{Larkoski:2014wba} and Dynamical Grooming (DyG)~\cite{Mehtar-Tani:2019rrk, Caucal:2021cfb} algorithms, developed in recent years, have been extensively employed in jet substructure studies from proton-proton to heavy-ion collisions~\cite{Chien:2016led, Mehtar-Tani:2016aco, Milhano:2017nzm, Caucal:2019uvr, Chang:2017gkt, Li:2017wwc, Ringer:2019rfk, Caucal:2021cfb, Cunqueiro:2022svx, Wang:2022yrp, JETSCAPE:2023hqn}. In this work, we will employ these two grooming algorithms to study the substructures of inclusive jets and $\gamma+$jets. Usually, the jet grooming procedure can be divided into three steps:

\begin{enumerate}
    \item Employing the anti-$k_T$ algorithm~\cite{Cacciari:2008gp} to recombine a large number of final-state particles into jets with a certain cone size $R$. Jets within specified $p_T$ and $\eta$ intervals are selected for analysis.
    \item Using the Cambridge/Aachen (C/A) algorithm~\cite{Dokshitzer:1997in} with constituents of the candidate jets to construct the clustering tree.
    \item Unwinding the C/A clustering tree to examine the subjet pair at each splitting vertex recursively until finding a pair satisfying the specific grooming condition.
\end{enumerate}

Note that SDG and DyG only differ in their criteria for identifying the hardest splitting during the third step of the above grooming procedure.

\textbf{Soft-Drop Grooming.} The SDG is a generalization and simplification of the modified mMDT algorithm~\cite{Dasgupta:2013ihk}. It operates with two parameters ($z_{\rm cut}, \beta$), which set an energy threshold and control the angular weighting for subjets, respectively.

\begin{eqnarray}
z>z_{\rm cut}(\frac{\Delta R_{12}}{R})^\beta,
\label{eq:sdg}
\end{eqnarray}
where $z=p_{T,2}/(p_{T,1}+p_{T,2})$ is the momentum fraction of the subleading subjet in the parent splitting, $p_{T,1}$ and $p_{T,2}$ are the transverse momentum of the leading and subleading subjets, respectively. $\Delta R_{12}=\sqrt{(\eta_1-\eta_2)^2+(\phi_1-\phi_2)^2}$ represents the angular distance between these two subjets in the $\eta-\phi$ plane and $R$ is the jet cone size. The Soft-Drop algorithm declusters a jet recursively according to the C/A clustering sequence, applying the condition specified in Eq.~(\ref{eq:sdg}) to the subjet pair. If the condition is satisfied, the two prongs of the subjet pair are identified as the hardest splitting; otherwise, the declustering procedure continues until a qualifying pair is found. Once a qualifying subjet pair is found in the C/A clustering sequence, we obtain the groomed momentum fraction $z_g=z$ and groomed radius $R_g=\Delta R_{12}$. It should be noted that for some jet events, no subjet pair satisfying the condition may be identified even after exhaustive declustering. Soft-Drop is infrared/collinear (IRC) safe only for $\beta \ge 0$. We set $\beta=0$ in this study when using the SDG algorithm.

\textbf{Dynamical Grooming.} Compared to the SDG, the DyG represents a distinct, dynamic approach that does not rely on a fixed threshold (e.g., $z_{\rm cut}$). Instead, it defines a ``hardness'' variable $\kappa$ and dynamically searches for the two subjets that maximize $\kappa$ while traversing the entire C/A clustering tree.

\begin{eqnarray}
\kappa = z(1-z)p_{\rm T, split}(\frac{\Delta R_{12}}{R})^a,
\label{eq:dyg}
\end{eqnarray}
 where the definitions of $z$ and $\Delta R_{12}$ are the same as in SDG, and $p_{\rm T, split}$ is transverse momentum of the parent splitting. The tunable parameter $a=0,1,2$ corresponds to the three distinct modes: $z$-Drop (select splittings with the most symmetric transverse momentum), $k_t$-Drop (chooses splittings with the largest relative transverse momentum), and TimeDrop (selects splittings with the shortest formation time), respectively~\cite{Mehtar-Tani:2019rrk}. The selected subjet pair is identified as the hardest splitting during the jet formation. Meanwhile, the groomed momentum fraction and groomed radius for DyG are determined as $z_g=z$ and $R_g=\Delta R_{12}$, respectively. Note that the DyG is an IRC-safe algorithm only for $a>0$.

\begin{figure}[!t]
\begin{center}
\includegraphics[width=2.9in,angle=0]{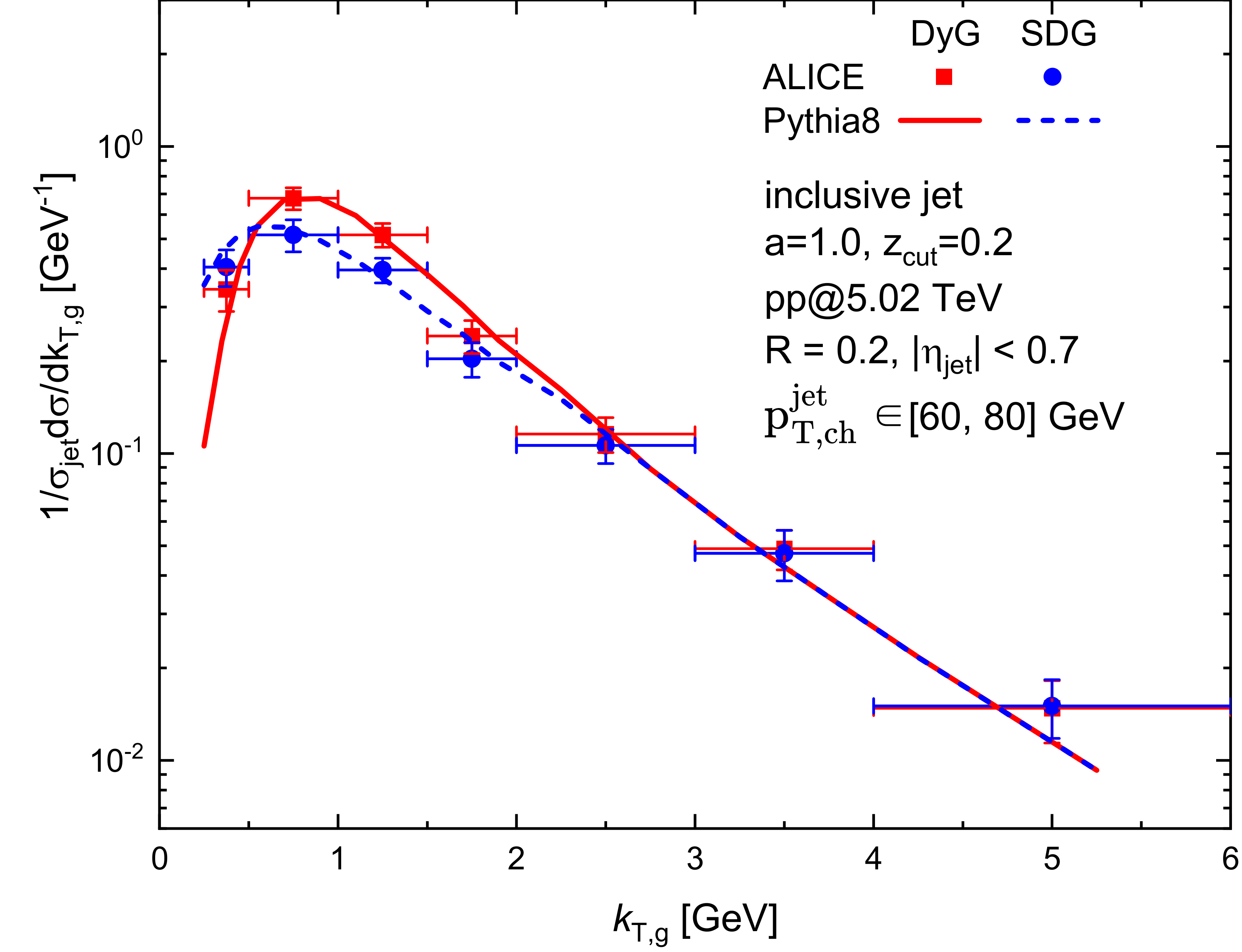}
\caption{(Color online) $k_{\rm T, g}$ distribution of the groomed inclusive jets in pp collisions at $\sqrt{s}=5.02$ TeV utilizing both the Dynamical Grooming ($a=1.0$) and Soft-Drop Grooming ($z_{\rm cut}=0.2$) algorithms, compared to the ALICE data~\cite{ALICE:2024fip}.}
\label{fig:alicepp}
\end{center}
\end{figure}

\begin{figure}[!t]
\begin{center}
\includegraphics[width=2.8in,angle=0]{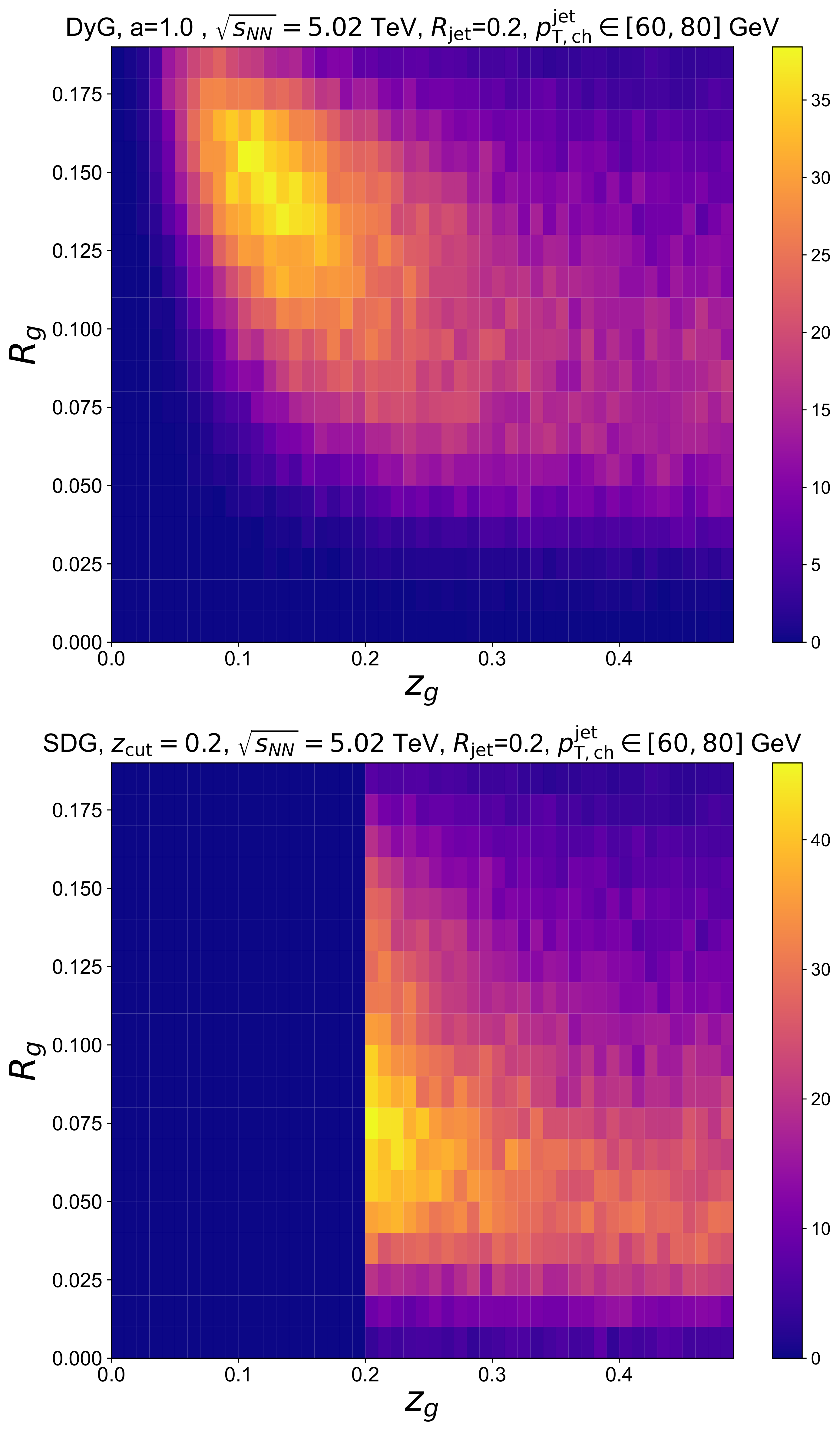}
\caption{(Color online) 2-dimensional ($z_g, R_g$) distribution of inclusive jets in pp collisions at $\sqrt{s}=5.02$ TeV computed using the Dynamical Grooming ($a=1.0$) and Soft-Drop Grooming ($z_{\rm cut}=0.2$) in the top and bottom figures.}
\label{fig:zgrg_pp}
\end{center}
\end{figure}

In Fig.~\ref{fig:alicepp}, we show the theoretical calculations of $k_{\rm T, g}$ distribution of the groomed inclusive jets in pp collisions at $\sqrt{s}=5.02$ TeV utilizing both the DyG ($a=1.0$) and SDG ($z_{\rm cut}=0.2$) algorithms, compared to the recent experimental data measured by the ALICE Collaboration~\cite{ALICE:2024fip}. Here $k_{\rm T, g}$ represents the relative transverse momentum between the two subjets,

\begin{eqnarray}
k_{\rm T, g}=p_{\rm T, 2}\cdot \sin \Delta R_{12}.
\label{eq:ktgdef}
\end{eqnarray}
Note that in this work we employ the Pythia8 \cite{Sjostrand:2014zea} event generator with the Monash Tune \cite{Skands:2014pea} to calculate the production of inclusive jets in pp collisions. The charged anti-$k_T$ jets with $R=0.2$ are required to have $60<p_T<80$ GeV and $|\eta|<0.7$. One can find that the Pythia8 calculations are in good agreement with the ALICE data in the measured $k_{\rm T, g}$ range. The $k_{\rm T, g}$ distributions calculated with the DyG and SDG are consistent at large $k_{\rm T, g}$ and only show a difference at $k_{\rm T, g}<2$ GeV.

To intuitively illustrate the differences of the jet substructures groomed by the DyG and SDG algorithms, in Fig.~\ref{fig:zgrg_pp} we show the 2-dimensional ($z_g, R_g$) distribution of inclusive jets in pp collisions at $\sqrt{s}=5.02$ TeV computed using the DyG ($a=1.0$) and SDG ($z_{\rm cut}=0.2$) in the top and bottom plots, respectively. By comparing the characteristics of the $z_g$ and $R_g$ distributions in the two figures, we observe that the jet substructures obtained using the SDG algorithm are predominantly concentrated in the region of $z_g \in$ [0.2, 0.3] and $R_g \in$ [0.05, 0.1]. Due to $z_{\rm cut}=0.2$, no splitting with $z_g < 0.2$ is retained. Conversely, DyG---lacking a $z_{\rm cut}$ constraint---retains substructures within $z_g\in$ [0.1, 0.2] and $R_g\in$ [0.125, 0.175]. This is because the DyG algorithm does not constrain the momentum of the subleading subjet, but instead maximizes the hardness variable $\kappa$. Through its adaptive selection strategy, DyG may permit jet substructures with larger angular distances and imbalanced momentum sharing. In contrast, SDG allows for manual control of soft radiation contributions through the setting of $z_{\rm cut}$.

Now we revisit the differences in the $k_{\rm T, g}$ distributions obtained with the two algorithms in the ALICE measurements as shown in Fig.~\ref{fig:alicepp}. The region of $k_{\rm T, g}\sim 0$ indicates that the two subjets are nearly collinear ($R_g \rightarrow 0$), where the probability for a splitting to satisfy the DyG selection criteria is significantly lower than for SDG. It can also be seen from Fig.~\ref{fig:zgrg_pp}: the $R_g$ distribution computed using SDG spans the entire [$0, R$] interval, while for DyG, it is almost zero for $R_g < 0.03$. Furthermore, in the intermediate $k_{\rm T, g}$ region, the DyG algorithm permits the splittings with larger radiation angles and smaller momentum fractions, leading to a higher yield for DyG in this region. Finally, since the high-$k_{\rm T, g}$ region corresponds to relatively hard splitting processes, these splittings naturally satisfy the selection criteria of both the DyG and SDG, resulting in identical distributions within this region.

The Pythia8 event generator used in this work provides a good description of the measured $k_{\rm T,g}$ distribution in pp collisions, including the high-$k_{\rm T,g}$ region. The agreement with ALICE data supports the use of this baseline for computing the nuclear modification ratio $\mathrm{PbPb}/\mathrm{pp}$. While higher-order corrections may affect the absolute $k_{\rm T,g}$ distribution, their impact is expected to remain under control in our analysis. This is because, as we have tested, the charged-jet $p_{T,{\rm ch}}^{\rm jet}>40$~GeV threshold employed in this study effectively selects events where the corresponding full jet $p_T$ predominantly lies in the range $80$--$140$~GeV, thereby strongly suppressing contributions from multi-jet events associated with higher-order processes. Therefore, the contribution from higher-order corrections should be moderate for the kinematic region of interest in this work.

\section{Theoretical framework of jet transport in QGP}
\label{sec:framework}

In this work, we utilize the SHELL model to simulate the evolution and medium modifications of jets within the QGP. In recent years, the SHELL model has been successfully applied to a series of theoretical studies concerning the production of heavy-flavor hadrons~\cite{Wang:2021jgm, Wang:2023udp}, as well as the medium modifications of full jets~\cite{Li:2024uzk, Wang:2020qwe, Wang:2024yag} and jet substructures~\cite{Wang:2019xey, Wang:2023eer, Wang:2020ukj, Li:2022tcr, Li:2024pfi, Wang:2024plm} in relativistic heavy-ion collisions. The current implementation of the SHELL model incorporates the effects of energy loss arising from both elastic and inelastic scattering between heavy/light partons and thermal constituents of the medium, along with the medium response due to the jet energy dissipation.

\textbf{Partonic jet energy loss:} The contribution of inelastic scattering, which dominates the partonic jet energy loss, is given by the medium-induced radiated gluon spectrum calculated within the higher-twist formalism~\cite{Guo:2000nz, Zhang:2003yn, Zhang:2003wk, Majumder:2009ge}.

\begin{eqnarray}
\frac{dN}{ dxdk^{2}_{\perp}dt}=\frac{2\alpha_{s}C_iP(x)\hat{q}}{\pi k^{4}_{\perp}}\sin^2(\frac{t-t_i}{2\tau_f}), \nonumber \\
\label{eq:dndxk}
\end{eqnarray}
 where $x$ and $k_{\perp}$ denote the energy fraction carried by the radiated gluon and its transverse momentum relative to the parent parton, respectively. $\alpha_{s}=0.3$ is the running coupling constant, and $C_i$ is the quadratic Casimir in color representation ($C_q=4/3$ for quarks and $C_g=3$ for gluons). $P(x)$ represents the QCD splitting function in vacuum, with primary consideration given to the dominant processes $q\rightarrow q+g$ and $g\rightarrow g+g$~\cite{Deng:2009ncl, He:2015pra}):

\begin{eqnarray}
P_{q\rightarrow q+g}(x)=&\frac{(1-x)(1+(1-x)^2)}{x},\\
P_{g\rightarrow g+g}(x)=&\frac{2(1-x+x^2)^3}{x(1-x)}.
\end{eqnarray}
$\tau_f=2x(1-x)E/k_{\perp}^2$ is the formation time of the radiated gluon. Due to the Landau-Pomeranchuk-Migdal (LPM) effect~\cite{Wang:1994fx, Zakharov:1996fv}, the emitted gluon can only undergo further interactions with the QGP after the formation time $\tau_f$. $\hat{q}=\hat{q}_0(T/T_0)^3p_{\mu}u^{\mu}/E$ is the jet transport parameter, characterizing the interaction strength between energetic partons and the QGP medium~\cite{Chen:2010te}, where $T_0$ is the medium temperature at the center of QGP with $\tau=0.6$ fm, $p_{\mu}$ and $u^{\mu}$ are the velocities of the parton and the medium cell. As a key parameter in the SHELL model, $\hat{q}_{0}=1.2 \pm 0.2$ GeV$^2$/fm has been extracted via a global $\chi^2$ fit to the yields of major identified hadron ($\pi^0$, $\eta$, $\rho^0$, $\phi$, $\omega$, and $K_{S}^0$) in PbPb collisions at the LHC~\cite{Ma:2018swx}. Furthermore, to implement the Monte Carlo simulation of the gluon radiation process, we assume radiation follows a Poisson distribution:

\begin{eqnarray}
P(n)=\frac{\lambda^ne^{-\lambda}}{n!}.
\end{eqnarray}
Here, $P(n)$ denotes the probability of $n$ radiation events occurring within a simulation time step $\Delta t=0.1$ fm. The parameter $\lambda$ of the Poisson distribution represents the average number of radiated gluons within $\Delta t$, obtained by integrating the spectrum of Eq.~(\ref{eq:dndxk}) over $t$, $x$, and $k_{\perp}$.

\begin{eqnarray}
\lambda (E, T, t) =\int_{t}^{t+\Delta t} dt\int_{x_{\rm min}}^{1} dx\int_{0}^{(xE)^2} dk^2_{\perp}\frac{dN}{dxdk^2_{\perp}dt}. \nonumber\\
\end{eqnarray}
Note that a lower cut $x_{\rm min}=\mu_D/E$ has been applied in the $x$ integral, where $\mu_D=\sqrt{6\pi\alpha_s}T$ is the Debye screening mass in the QGP medium. In each simulation step $\Delta t$, we first calculate the total radiation probability $P_{\rm rad}=1-P(0)$ to determine whether radiation occurs. Once radiation is triggered, the number of gluons is determined by the Poisson distribution, and the energy-momentum of each gluon is sequentially sampled stochastically according to the higher-twist radiated gluon spectrum. Additionally, the contribution of elastic energy loss of the jet in the QGP is provided by a pQCD calculation based on the hard-thermal-loop approximation~\cite{Braaten:1991we, Neufeld:2010xi},

\begin{eqnarray}
\frac{dE}{dL}=-\frac{8\pi\alpha_s^2T^2}{3}(1+\frac{N_f}{6}){\rm ln}{\frac{q_{\rm max}}{q_{\rm min}}},
\label{eq:HTL}
\end{eqnarray}
where $L$ represents the path length along the parton's momentum direction. $q_{\rm max}=\sqrt{ET}$ and $q_{\rm min}=\mu_D$ are the upper and lower limits for the 3-momentum transfer between the jet parton and the medium. During each time step $\Delta t$, the collisional energy loss of a parton can be calculated by integrating Eq.~(\ref{eq:HTL}).

In this work, we use complete pp events generated by Pythia8 as input for the SHELL simulation of nucleus-nucleus collisions to study medium modifications of jet substructure. The spatial vertex of hard scatterings is sampled using the MC-Glauber model~\cite{Miller:2007ri}. Simultaneously, the space-time evolution of the hot and dense nuclear matter produced in the collisions is described by the CLVisc hydrodynamic model~\cite{Pang:2016igs}, which provides information such as the local temperature and flow velocity of the medium at the jet parton's location. Jet partons stop interacting with the medium when the local temperature drops below $T_c = 165$ MeV; then hadronization occurs. The SHELL model adopts the Colorless Hadronization Scheme developed by the JETSCAPE collaboration~\cite{Putschke:2019yrg}, which is based on the Lund string fragmentation model~\cite{Andersson:1983jt, Sjostrand:1984ic}.

\textbf{Medium response:} The SHELL model also incorporates medium response effects, which are essential for studying the medium modifications of the internal structure of jets in nucleus-nucleus collisions. During the jet-QGP interaction, quasiparticles within the medium are excited and become part of the jet, thereby influencing jet substructure observables. This medium response effect can be implemented via a modified Cooper-Frye formula that accounts for perturbations on the velocity ($\delta u^{\mu}$) and temperature ($\delta T$) of the expanding medium caused by the energy dissipation from energetic jets~\cite{Cooper:1974mv}. The momentum spectra of particles hadronized from the QGP can be expressed as:

\begin{eqnarray}
E\frac{dN}{d^3p}=\frac{1}{(2 \pi)^3}\int d\sigma^{\mu}p_{\mu}f(u^{\mu}p_{\mu}, \delta u^{\mu}, \delta T),
\label{eq:cooper}
\end{eqnarray}
where the $\sigma^{\mu}$ integral is over the freeze-out hypersurface and $p_{\mu}$ is the four-momentum of particles. The perturbed phase-space distribution function $f(u^{\mu}p_{\mu}, \delta u^{\mu}, \delta T)$ can be expressed through a first-order Taylor expansion of the Boltzmann distribution $f(E)=\exp[-E/T]$ for $\delta u^{\mu}$ and $\delta T$:

\begin{eqnarray}
f(u^{\mu}p_{\mu}, \delta u^{\mu}, \delta T)=\exp[-\frac{u^{\mu}p_{\mu}}{T}+\frac{u^{\mu}p_{\mu}}{T^2}\delta T+\frac{p_{\perp}^i}{T}\delta u_{\perp}^i].\nonumber\\
\label{eq:pertf}
\end{eqnarray}

By relating the perturbations of velocity and temperature to the total energy dissipation from the traversing jets, the momentum spectra of quasiparticles excited from the medium can be approximately obtained as follows, and more details can be found in Ref.~\cite{Casalderrey-Solana:2016jvj}.

\begin{eqnarray}
E\frac{d\Delta N}{d^3p}&=&\frac{m_T}{32\pi T^5}{\rm cosh(\Delta y)\exp}[-\frac{m_T}{T}{\rm cosh(\Delta y)]} \nonumber\\
&&\times \{p_T\Delta P_{T}{\rm cos(\Delta \phi)}+\frac{1}{3}m_T\Delta M_T{\rm  cosh(\Delta y)}\}, \nonumber\\
\label{eq:resp}
\end{eqnarray}
where $\Delta y=y-y_{\rm jet}$ and $\Delta \phi=\phi-\phi_{\rm jet}$ represent the differences in rapidity and azimuthal angle between the excited quasiparticle and the jet axis, respectively. $m_T$ and $p_T$ are the transverse mass and transverse momentum of the particle. $\Delta P_T$ and $\Delta M_T = \Delta E / \cosh y_{\rm jet}$ are the changes in transverse momentum and transverse mass of jets during the in-medium evolution, where $\Delta E$ is the jet energy loss.  After the jet completes its evolution in the medium, the $\Delta P_T$ and $\Delta E$ of the jet can be obtained. Then, based on  Eq.~(\ref{eq:resp}), we perform Monte Carlo sampling for $p_T$, $\phi$, and $y$ of the excited quasiparticles. During the sampling, the value of $\Delta E$ is reduced by the energy of the excited quasiparticles accordingly until $\Delta E=0$. At this point, all energy lost by the jet is considered dissipated into the QGP in the form of excited quasiparticles to guarantee energy conservation.

\begin{figure*}[!t]
\begin{center}
\includegraphics[width=4.8in,angle=0]{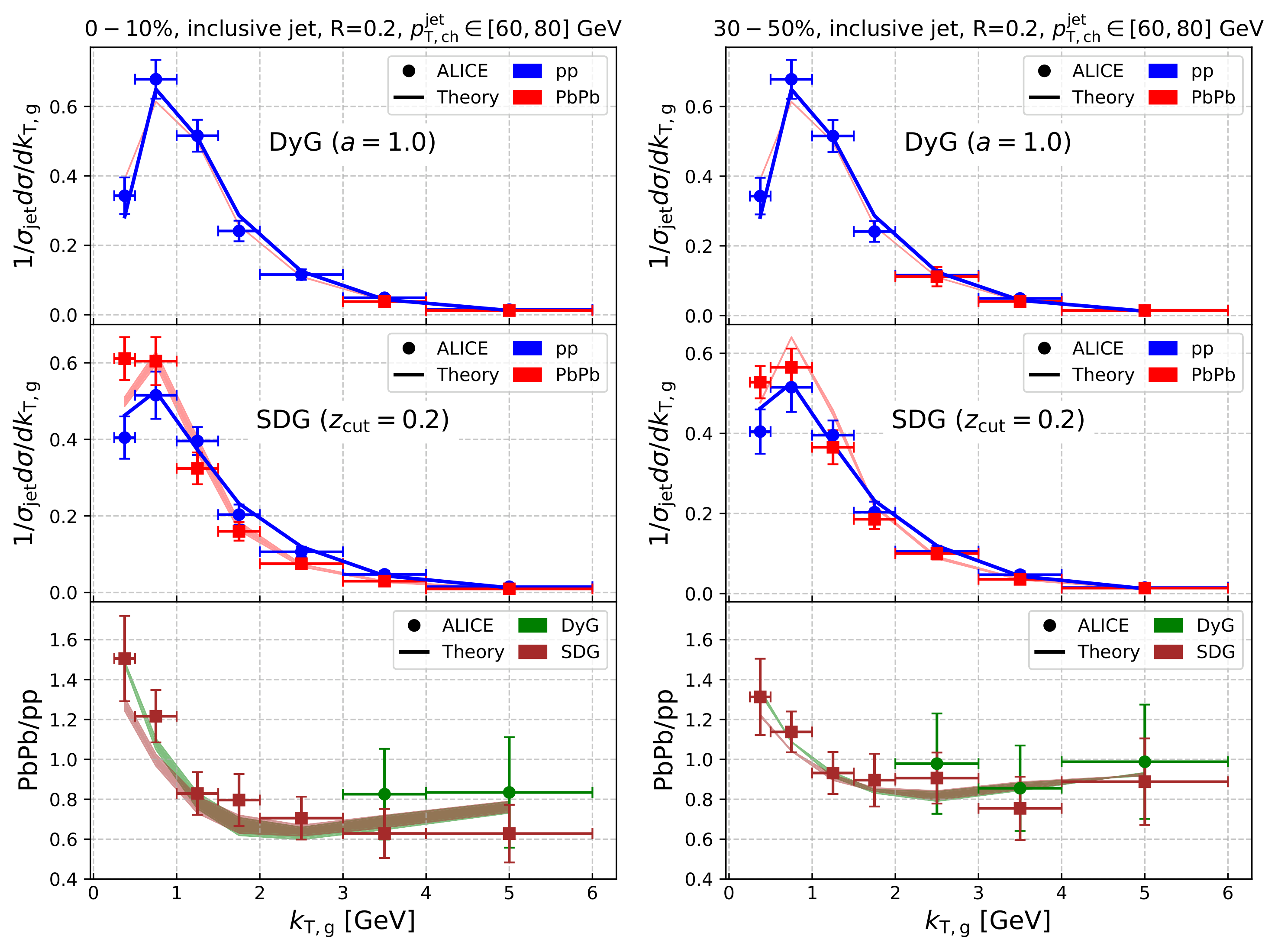}
\caption{(Color online) $k_{\rm T, g}$ distributions of the groomed inclusive jets in pp and PbPb (left panel: $0-10\%$, right panel: $30-50\%$) collisions at $\sqrt{s_{NN}}=5.02$ TeV utilizing both the Dynamical Grooming (top panels) and Soft-Drop Grooming (middle panels) algorithms, compared to the recent experimental data measured by the ALICE Collaboration~\cite{ALICE:2024fip}. Each distribution is normalized by the total jet cross section in the $p_T$ interval $[60, 80]$ GeV. The ratios of PbPb/pp are also shown in the bottom panels.}
\label{fig:alice-ktg}
\end{center}
\end{figure*}

\section{Results and discussions}
\label{sec:res}

\begin{figure*}[!t]
\begin{center}
\includegraphics[width=4.6in,angle=0]{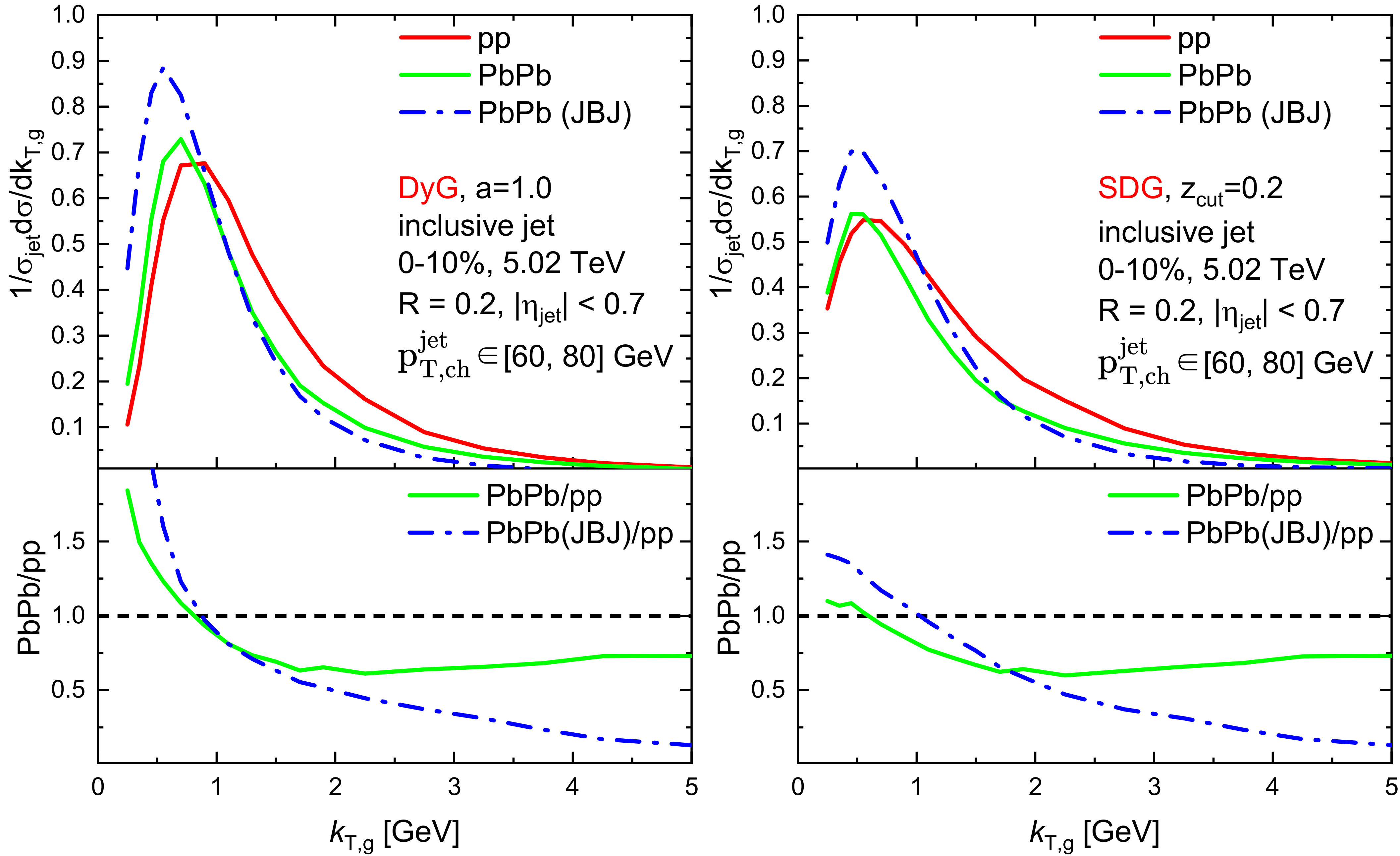}
\vspace*{-0in}
\caption{(Color online) $k_{\rm T, g}$ distributions of the groomed inclusive jets in pp and $0-10\%$ PbPb collisions at $\sqrt{s_{NN}}=5.02$ TeV utilizing both the Dynamical Grooming (left plot) and Soft-Drop Grooming (right plot) algorithms. The calculations by the JBJ matching analysis are also shown for comparison. The ratios of PbPb/pp are also shown in the bottom panels.}
\label{fig:incl-ktg}
\end{center}
\end{figure*}

\begin{figure*}[!t]
\begin{center}
\includegraphics[width=4.6in,angle=0]{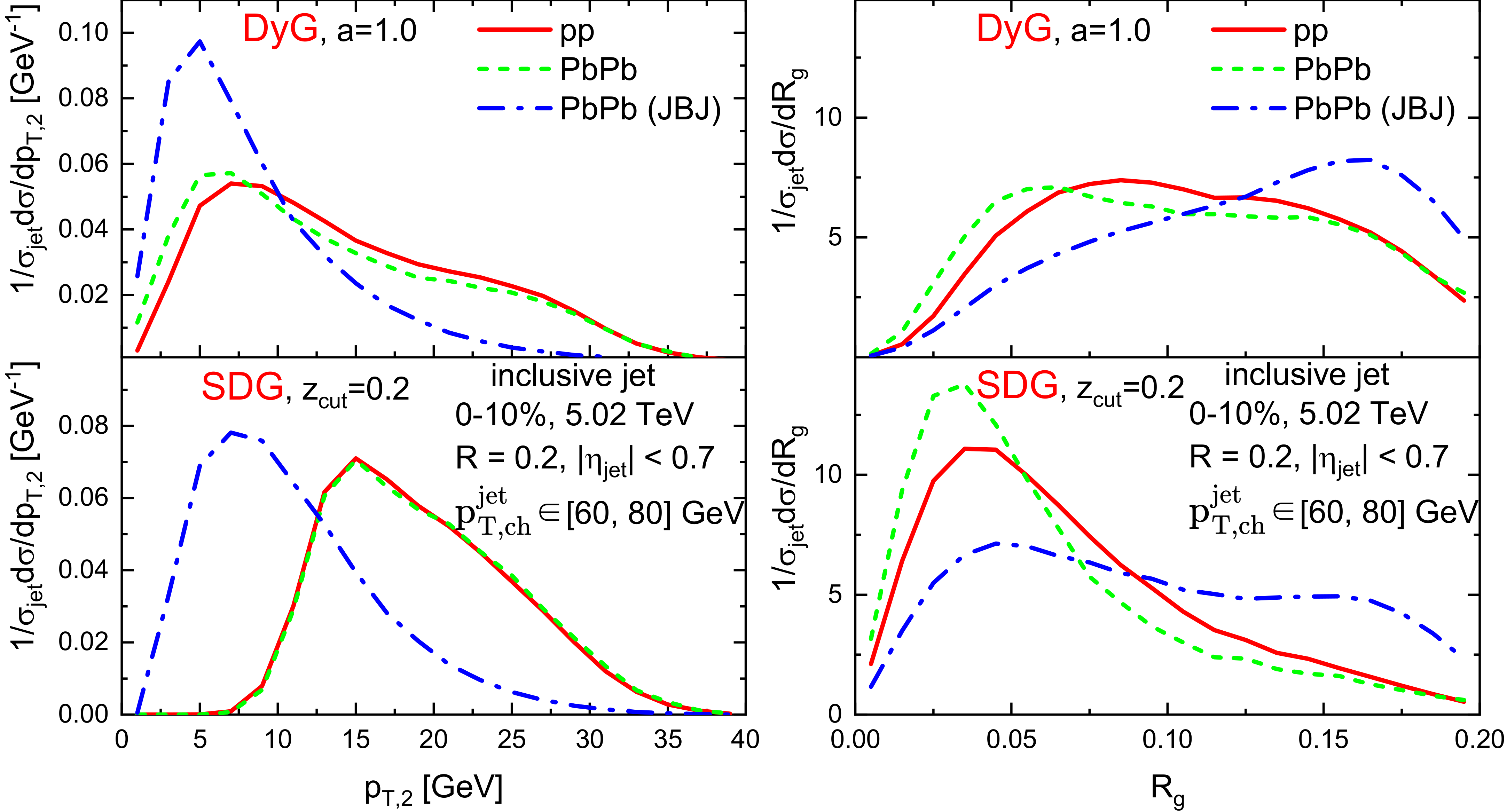}
\vspace*{0in}
\caption{(Color online) Normalized distributions of $p_{T,2}$ and $R_g$ of the groomed inclusive jets in pp and $0-10\%$ PbPb collisions at $\sqrt{s_{NN}}=5.02$ TeV utilizing both the Dynamical Grooming (left plot) and Soft-Drop Grooming (right plot) algorithms. The calculations by the JBJ matching analysis are also shown for comparison.}
\label{fig:incl-pt}
\end{center}
\end{figure*}

In Fig.~\ref{fig:alice-ktg}, we present the theoretical calculations of the groomed splitting momentum $k_{\rm T,g}$ of inclusive jets in $0-10\%$ and $30-50\%$ PbPb collisions as compared to pp at $\sqrt{s_{NN}}=5.02$ TeV. The top and middle panels display the $k_{\rm T,g}$ spectra computed using the DyG and SDG algorithms, respectively, while the bottom panel shows the ratio of the $k_{\rm T,g}$ spectra in PbPb collisions relative to the pp baseline. A total of 10 million inclusive jet events are analyzed in the calculations. By comparing with experimental data at two centrality classes~\cite{ALICE:2024fip}, the model calculations provide a reasonable description of the $k_{\rm T,g}$ spectra and their medium modifications. Specifically, compared to pp, the $k_{\rm T,g}$ spectra in PbPb collisions exhibit a characteristic enhancement at smaller values and a suppression at larger values. Furthermore, by comparing the results from the two jet grooming algorithms, we observe that for the PbPb/pp ratio of $k_{\rm T,g}$, the calculation using the DyG algorithm yields a larger modification than that using the SDG algorithm at $k_{\rm T,g}< 1$ GeV. No significant difference between the two algorithms is found at high $k_{\rm T,g}$. The suppression of $k_{\rm T,g}$ at high transverse momentum indicates that the relative transverse momentum between the two subjets in PbPb collisions is reduced. This apparent inconsistency with the conventional picture of jet-medium interactions arises because scattering between jet partons and quasiparticles in the QGP typically leads to a broadening, rather than a narrowing, of the jet substructure. Note that the error bands of the model calculation in PbPb collisions shown in Fig.~\ref{fig:alice-ktg} arise from the uncertainties in the model parameter $\hat{q}_{0}$. Since the $k_{\rm T,g}$ modification seems not particularly sensitive to the variation of $\hat{q}_{0}$, we will use its central value in subsequent calculations and discussions for simplicity.

The narrowing of the $k_{\rm T,g}$ distribution of inclusive jets was attributed primarily to selection bias, as suggested by the ALICE Collaboration~\cite{ALICE:2024fip}, a phenomenon that has also been observed in other jet measurements~\cite{ALargeIonColliderExperiment:2021mqf, ATLAS:2022vii, ALICE:2018dxf, ALICE:2023dwg, Ehlers:2022dfp, ATLAS:2023hso}.
Selection bias refers to the scenario where, despite applying an identical selection $p_T$ threshold to jet identification in both pp and PbPb collisions, the dynamical ranges represented by the selected jet samples are mismatched. To ascertain whether the narrowing of $k_{\rm T,g}$ stems from the contamination by selection bias similar to other observables~\cite{ALargeIonColliderExperiment:2021mqf, ATLAS:2022vii, ALICE:2018dxf, ALICE:2023dwg, Ehlers:2022dfp, ATLAS:2023hso}, in the following we employ a jet-by-jet (JBJ) matching analysis to make a detailed investigation~\cite{Brewer:2021hmh, Kang:2023ycg}.

The JBJ matching method involves matching jets before and after quenching and directly comparing them. The JBJ matching method minimizes the selection bias and reveals the true modifications of the same jet before and after quenching. The specific procedure is as follows:
\begin{enumerate}
    \item Jet reconstruction is first performed on pp events before medium evolution, and the pre-quenching jet information (e.g., $p_T$, $\phi$, $\eta$) is recorded.
    \item After the medium evolution of the pp events is complete, jet reconstruction is carried out on the quenched events using a lower $p_T$ threshold (e.g., 10 GeV).
    \item Subsequently, each jet ($i$) reconstructed in PbPb collisions is matched to its counterpart ($j$) from the pp events if the angular separation between them in the $\eta-\phi$ plane satisfies $\Delta R_{ij} < R$, thereby identifying the PbPb jet as the quenched version of the pp jet.
\end{enumerate}
Note that employing a lower $p_T$ threshold for jet reconstruction in the PbPb events helps prevent jets from being discarded due to energy loss, thereby improving the matching efficiency. In actual experiments, it is not easy to achieve this, but it is feasible in Monte Carlo simulations. The JBJ matching method effectively circumvents the influence of selection bias and focuses on the real modifications of a jet induced by jet-medium interactions.

In Fig.~\ref{fig:incl-ktg}, we present a comparison between the $k_{\rm T,g}$ results obtained using the jet-by-jet matching analysis method and those from the default calculation. The bottom panel also shows the corresponding PbPb/pp ratios. As one can see, the JBJ results show stronger suppression than the default calculation at high $k_{\rm T,g}$. It indicates that if the selection bias is eliminated, we will observe a more pronounced narrowing of the $k_{\rm T,g}$ distribution. This finding suggests that the narrowing of $k_{\rm T,g}$ observed in PbPb collisions is not solely attributable to selection bias. This characteristic distinguishes it from other observables, such as girth~\cite{Wang:2024plm}. To elucidate this finding, we will separately discuss the nuclear modification behavior of the momentum component $p_{T,2}$ and angular component $R_g$ of $k_{\rm T,g}$ in PbPb collisions.

Fig.~\ref{fig:incl-pt} displays the distributions of $p_{T,2}$ and $R_g$ calculated using both the DyG and SDG algorithms in pp and PbPb collisions, alongside the JBJ results for comparison. We observe in the left figure that when using the DyG algorithm, the $p_{T,2}$ distribution in PbPb collisions exhibits a shift towards smaller values compared to pp collisions. This trend is attributed to jet energy loss resulting from interactions between jet partons and the QGP medium. In contrast, when the SDG algorithm is employed, the $p_{T,2}$ distribution in PbPb collisions remains almost unchanged relative to the pp baseline. This coincidence between pp and PbPb arises from the combined effect of the global jet selection criterion ($60<p_T<80$ GeV) and the Soft-Drop condition ($z_{\rm cut}=0.2$). Under these conditions, subjets that have undergone significant energy loss in PbPb collisions are removed by grooming, leading to similar $p_{T,2}$ distributions in PbPb and pp collisions. However, this apparent consistency does not imply an absence of energy loss for the subjets; instead, it should be interpreted as an additional bias introduced by the event selection criteria. The influence of this selection bias can be effectively eliminated by applying a JBJ matching analysis, which tracks the medium modifications of individual jets. As shown in the figure, the decreasing trend of $p_{T,2}$ becomes more pronounced in the JBJ matching analysis for both the DyG and SDG algorithms.

Furthermore, in the right plot of Fig.~\ref{fig:incl-pt}, we apply the same JBJ matching analysis to the groomed radius $R_g$. We find that the $R_g$ distribution of inclusive jets in PbPb collisions shows a significant narrowing trend compared to pp collisions. As detailed in our previous work~\cite{Wang:2024plm}, this jet narrowing, contrary to the expected broadening, is primarily due to selection bias. Notably, we observe that when the JBJ matching analysis is employed, the $R_g$ distribution in PbPb collisions exhibits significant broadening, in contrast to the narrowing trend of the default PbPb/pp results. Since the JBJ matching analysis is free from selection bias, the observed broadening provides direct evidence of the impact of jet-parton interactions with the medium on the jet substructure.

\begin{figure*}[!t]
\begin{center}
\includegraphics[width=4.6in,angle=0]{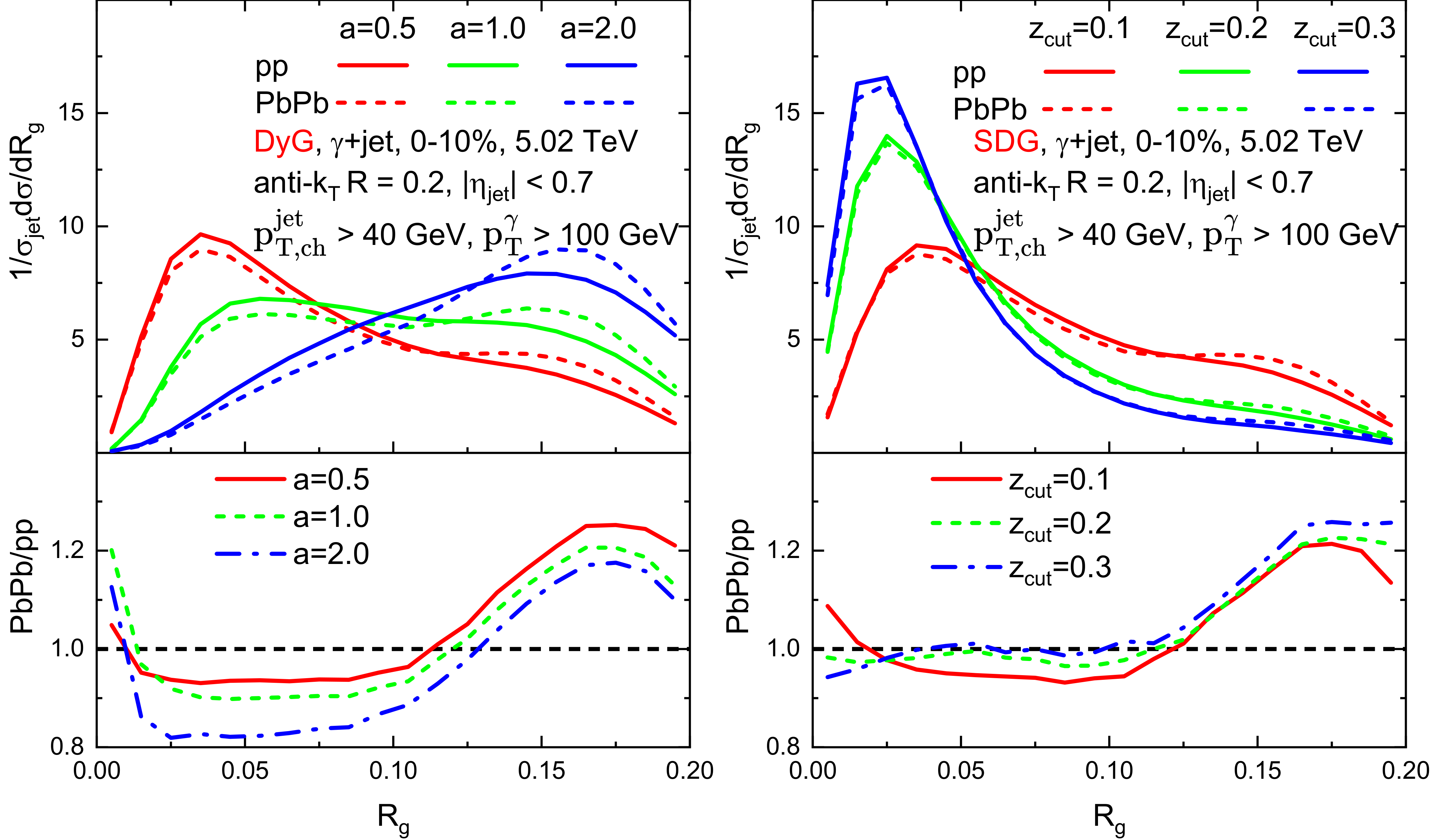}
\vspace*{0in}
\caption{(Color online) Normalized $R_g$ distributions of the groomed $\gamma+$jets in pp and $0-10\%$ PbPb collisions at $\sqrt{s_{NN}}=5.02$ TeV utilizing both the Dynamical Grooming (left plot) and Soft-Drop Grooming (right plot) algorithms. The calculations with different grooming parameters ($a=0.5$, $1$, $2$ and $z_{\rm cut}=0.1$, $0.2$, $0.3$) are shown for comparison.}
\label{fig:gi-dR-para}
\end{center}
\end{figure*}

\begin{figure*}[!t]
\begin{center}
\includegraphics[width=4.6in,angle=0]{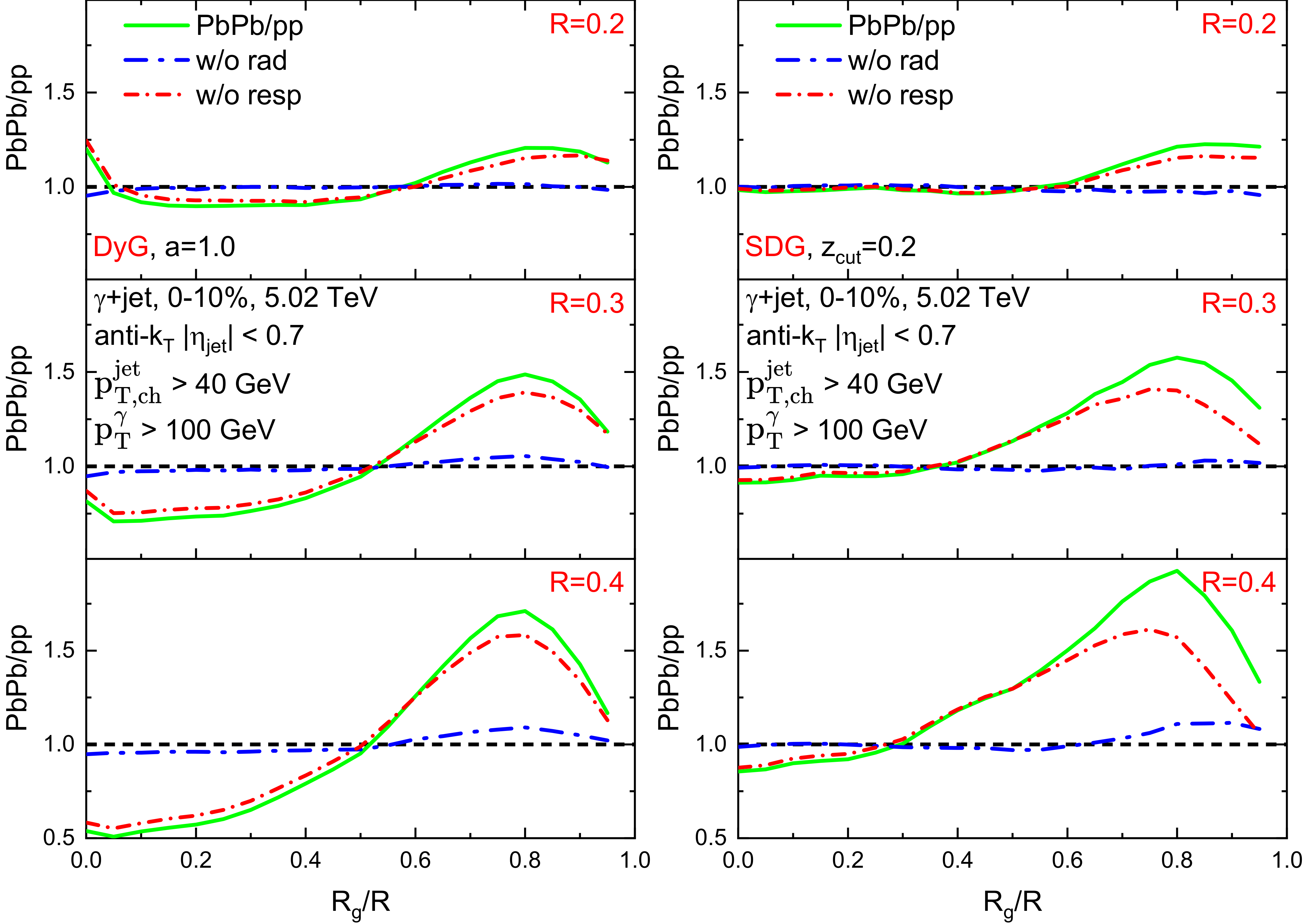}
\caption{(Color online) Medium modification of $R_g$ distributions of the groomed $\gamma+$jets in $0-10\%$ PbPb collisions compared to pp at $\sqrt{s_{NN}}=5.02$ TeV are calculated with different jet radius ($R=$0.2, 0.3, 0.4) in the top to bottom panels utilizing both the Dynamical Grooming (left plot) and Soft-Drop Grooming (right plot) algorithms. The calculations, excluding medium-induced gluon radiation and medium response, are also shown for comparison.}
\label{fig:gi-dR-R}
\end{center}
\end{figure*}

Analysis of $p_{T,2}$ and $R_g$ of inclusive jets reveals that energy loss decreases $p_{T,2}$ in PbPb collisions, while selection bias causes a narrowing of $R_g$. The combination of these effects makes it highly unlikely to observe an enhancement at high $k_{\rm T,g}$ since $k_{\rm T, g}= p_{T, 2}\cdot \sin R_g$. This elucidates the underlying reason for the suppression observed at high $k_{\rm T,g}$ in the ALICE measurements, which differs from the narrowing of $\Delta R_{\rm axis}$ and girth and cannot be simply attributed to selection bias. Meanwhile, we recognize that even in the JBJ matching analysis, which is utterly devoid of selection bias, although $R_g$ exhibits broadening, $p_{T,2}$ is reduced more significantly. Consequently, the net effect is a narrowing of $k_{\rm T,g}$, as shown in Fig.~\ref{fig:incl-ktg}. In other words, $k_{\rm T,g}$  is not suitable for probing the jet broadening effect induced by jet scattering with quasiparticles.

From the above analysis, we find that if the influence of selection bias can be eliminated or reduced, it is promising to search for jet broadening utilizing the observable $R_g$. Fortunately, it can be achieved with photon-tagged jets ($\gamma$+jets), which substantially mitigate selection bias effects compared to inclusive jets. This method has been demonstrated to be feasible in the measurement of jet girth by CMS~\cite{CMS:2024zjn}. In Ref.~\cite{Wang:2024plm}, this issue has been systematically and quantitatively investigated, where we demonstrated for the first time that, compared to inclusive jets, $\gamma$+jets not only effectively suppress selection bias but also efficiently select those jets that have undergone significant quenching. Consequently, in the remainder of this paper, we will focus on the systematic theoretical investigations and predictions for the $R_g$ modification of $\gamma$+jets in nucleus-nucleus collisions. This study will comprehensively examine the $R_g$ modification under various grooming algorithms, parameter settings, and jet radii, and assess its sensitivity to jet-medium interaction mechanisms and hadronization processes. It should be noted that in the subsequent analysis, we will no longer discuss the $k_{\rm T,g}$ of $\gamma$+jets, but will instead focus on investigating the nuclear modification of $R_g$. This is because simply removing or mitigating the selection bias has been shown to be insufficient for producing a broadened $k_{\rm T,g}$ distribution, as illustrated in Fig.~\ref{fig:incl-ktg}. As supplementary material for this paper, the calculated results and a discussion concerning the nuclear modification of the $k_{\rm T,g}$ for photon-tagged jets are still provided in Appendix~\ref{sec:gjet-ktg}. Consistent with expectations, a narrowing of the $k_{\rm T,g}$ distribution is observed.

\begin{figure*}[!t]
\begin{center}
\includegraphics[width=4.6in,angle=0]{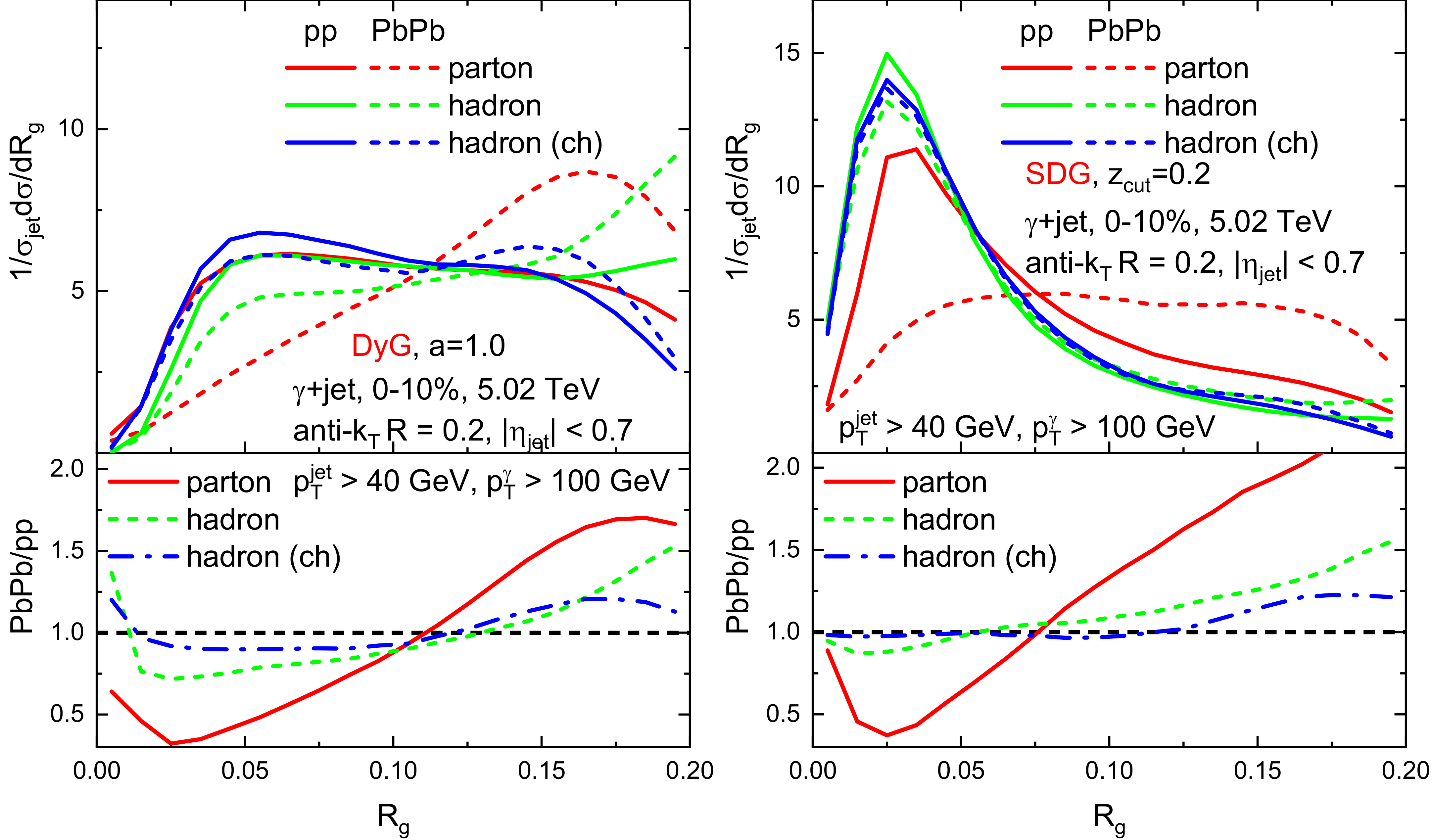}
\caption{(Color online) Normalized $R_g$ distributions of the groomed $\gamma+$jets in pp and $0-10\%$ PbPb collisions at $\sqrt{s_{NN}}=5.02$ TeV utilizing both the Dynamical Grooming (left plot) and Soft-Drop Grooming (right plot) algorithms. The calculations at the parton level and hadron level are shown for comparison.}
\label{fig:gi-dR-had}
\end{center}
\end{figure*}

As shown in Fig.~\ref{fig:gi-dR-para}, we first examine the influence of grooming algorithm parameters on the $R_g$ distributions of $\gamma$+jets in both pp and PbPb collisions.
Note that the $\gamma+$jets must satisfy the following criteria: the jets are reconstructed from charged hadrons with a threshold $p_{T,{\rm ch}}^{\rm jet}>40$~GeV and within a rapidity window $|\eta_{\rm jet}|<0.7$. The photon is required to have $p_T^{\gamma}>100$ GeV and must meet the isolation condition, namely that the sum of the energies of all particles within a cone of radius R=0.4 around the photon is less than the isolation cut on transverse energy $E_{\rm T, cut}^{\rm iso}=0.0042\cdot E_T^{\gamma}+4.8$ GeV~\cite{ATLAS:2019iaa}. Furthermore, the photon and the jet are required to be approximately a ``back-to-back'' configuration: $|\phi_{\rm jet}-\phi_{\gamma}|>2\pi/3$, in line with the CMS measurements~\cite{CMS:2024zjn}. Additionally, we set $a=0.5$, 1.0, and 2.0 for the DyG algorithm. For the SDG algorithm, which has two parameters ($z_{\rm cut}, \beta$), we focus on the variation of $z_{\rm cut}$ with values of 0.1, 0.2, and 0.3, while keeping $\beta=0$ fixed. A total of $10$ million $\gamma+$jet events are simulated and analyzed in the calculations. In the plotted results, we observe that increasing parameter $a$ in DyG shifts the $R_g$ distribution toward larger values. Since $a$ governs the weight assigned to the angular separation in the calculation of the splitting hardness, a larger $a$ favors splitting processes with larger opening angles. Regarding the nuclear modification of $R_g$, we note that as $a$ increases, the absolute difference in $R_g$ between PbPb and pp collisions becomes more pronounced. Meanwhile, because a larger value of $a$ also increases the pp distribution at larger $R_g$, the overall enhancement of the PbPb/pp ratio at larger angles is consequently attenuated. In the right panel, for the SDG algorithm, higher $z_{\rm cut}$ values exclude soft wide-angle splittings, narrowing the $R_g$ distribution both for pp and PbPb. We observe that the variation of $z_{\rm cut}$ has a less significant impact on the PbPb/pp ratio of $R_g$ compared to that of $a$ in the DyG. Most importantly, compared to pp collisions, the groomed jet radius $R_g$ distributions in PbPb collisions exhibit a distinct shift towards larger values across various grooming algorithms and parameter settings. This is accompanied by a pronounced suppression of the PbPb/pp ratio at small $R_g$ and an enhancement at large $R_g$. It indicates that, even without relying on the jet-by-jet matching, we can observe the jet substructure broadening in actual experimental measurements, in contrast to the narrowing of inclusive jets shown in Fig.~\ref{fig:incl-pt}.

We now proceed to investigate the contributions of different jet-medium interaction mechanisms to the broadening of $\gamma$+jet $R_g$ in PbPb collisions, as well as the dependence of this broadening on the jet radius $R$. In Fig.~\ref{fig:gi-dR-R}, the left and right panels present the calculated PbPb/pp ratio for $R_g$ using DyG and SDG, respectively. The three panels from top to bottom correspond to jet radii $R=0.2$, 0.3, and 0.4. Since the calculations involve different jet radii, to facilitate comparison, we present the distributions as a function of the scaled variable $R_g/R$. In addition to the default results for PbPb/pp, we also show calculations obtained by deactivating specific mechanisms within the SHELL model: the medium-induced gluon radiation and the medium response. Our analysis reveals two key findings. First, as the jet radius $R$ increases, the nuclear modification effects on $R_g$ become more pronounced. One can observe a more pronounced suppression of the PbPb/pp ratio at small $R_g/R$ and a more significant enhancement at large $R_g/R$. Second, the relative contributions of the underlying physical mechanisms exhibit a clear dependence on the jet radius. For jets with a small radius (e.g., R=0.2), the calculated $R_g$ distribution shows almost no observable broadening when the contribution of the medium-induced gluon radiation is disabled. It indicates that medium-induced radiation is the dominant mechanism responsible for the $R_g$ broadening in small-$R$ jets. The physical interpretation is that jets radiate gluons through inelastic scattering with quasiparticles in the QGP medium, leading to a redistribution of the jet energy and a widening of the opening angle between subjets.

Furthermore, as the jet radius increases, the influence of the medium response gradually becomes more significant. This effect is particularly notable when using the SDG algorithm, where the medium response plays a significant role in the medium modification of $R_g$ in the large $R_g/R$ region. It implies that while both medium-induced gluon radiation and medium response are, in essence, processes of energy dissipation from the jet into the medium, different grooming algorithms may show distinct sensitivities to these two mechanisms. More in-depth studies on this issue will likely be necessary in the future. Note that the $R$-dependence of the relative contributions arises from the distinct angular scales of the two mechanisms. Medium-induced radiation, characterized by a relatively small emission angle, remains largely within the jet core even for small $R$. In contrast, the medium response produces a wide-angle wake; a significant fraction of it falls outside the jet cone for small $R$, but is progressively captured as $R$ increases, enhancing its relative role in $R_g$ broadening.

Fig.~\ref{fig:gi-dR-had} presents the calculated $R_g$ distributions of $\gamma+$jets at both parton level and hadron level, aiming to investigate the sensitivity of different algorithms to hadronization effects in pp and PbPb collisions. First, in pp collisions, the difference between the $R_g$ distributions computed with the DyG algorithm at the parton level and the hadron level is smaller than that for the SDG algorithm. It indicates that the SDG algorithm may be more sensitive to hadronization effects than the DyG, which may be attributed to its reliance on a fixed $z_{\rm cut}$ threshold for selecting the hardest splitting process. Furthermore, in PbPb collisions, the parton-level $R_g$ distributions exhibit a significant broadening for both grooming algorithms. Due to the lower initial $R_g$ distribution in the large-$R_g$ region obtained with the SDG algorithm, the resulting nuclear modification (PbPb/pp) is more pronounced than that for DyG. However, we observe reduced magnitudes of the PbPb/pp ratio at the hadron level compared to the parton level, indicating the $R_g$ broadening is noticeably attenuated by the hadronization effect. Additionally, we find that reconstructing jets using only charged particles further weakens the observed $R_g$ broadening effect. These results underscore the necessity of appropriately accounting for hadronization effects in theoretical calculations to achieve a more accurate description of experimental data. Future high-precision measurements of jet substructure will provide essential constraints on the mechanisms of jet hadronization within the QGP.

A recent study employing the JETSCAPE framework investigates similar groomed observables for photon-tagged jets and attributes the observed non-monotonic modifications primarily to the dominant effect of recoil medium response~\cite{Tachibana:2025rcx}. Our work provides a complementary and distinct perspective. We offer a mechanistic decomposition of the high-$k_{\rm T,g}$ suppression and focus on proposing $R_g$ as a new, sensitive observable, providing systematic predictions for its broadening. The convergence on the utility of photon-tagged jets from different theoretical frameworks strengthens the case for this channel, while the differences in detailed interpretation enrich the understanding of jet-medium interactions. In addition, it is noteworthy that the kinematic selections for photon-tagged jets ($p_T^{\gamma} > 100$ GeV, $p_{T, {\rm ch}}^{\text{jet}} > 40$ GeV) in this work are adopted from recent CMS measurements~\cite{CMS:2024zjn}, ensuring that our theoretical predictions are directly comparable to experimental data. Given that the present study is based on the LO+PS Pythia8 calculation, a future extension of our framework to next-to-leading order matched with parton showers (NLO+PS) would be valuable to systematically quantify the impact of higher-order corrections on groomed jet substructure.

\section{Summary}
\label{sec:sum}

Grooming techniques mitigate non-perturbative effects in jets, providing a unique opportunity to gain insights into partonic interactions within the QGP at a finer scale. In this paper, we conduct a systematic theoretical study of the groomed substructures for both inclusive jets and $\gamma+$jets in pp and PbPb collisions, employing the DyG and SDG algorithms. In PbPb collisions, the jet--medium interactions are simulated using the SHELL model, which incorporates both partonic energy loss and medium response effects in the QGP.

At first, in pp collisions, we analyze the characteristics of two grooming algorithms, DyG and SDG, and explain the differences observed in the recent $k_{\rm T,g}$ measurements of inclusive jets by the ALICE Collaboration. Our calculations show a narrowing of the splitting momentum for inclusive jets in PbPb collisions compared to the pp baseline, in agreement with the ALICE data. Through a jet-by-jet matching analysis, we reveal that the suppression of high $k_{\rm T,g}$ arises from the combined effect of the reduction of $p_{T,2}$ due to partonic energy loss and the narrowing of $R_g$ induced by selection bias. Our findings indicate that no enhancement can be observed at high $k_{\rm T,g}$, even in the complete absence of selection bias.

Furthermore, we propose that the broadening of $R_g$ in photon-tagged jets---which are less susceptible to selection bias compared to inclusive jets---can provide more direct evidence of jet-medium interactions. Our analysis reveals that medium-induced gluon radiation plays a dominant role in driving the $R_g$ broadening. This broadening becomes more pronounced as the jet radius increases. The contribution from the medium response is particularly significant for large-$R$ jets when SDG is employed. Additionally, through the comparison of $R_g$ distributions at the parton and hadron level, we confirm that it is essential to include hadronization effects when studying the groomed substructure in nucleus-nucleus collisions. Against the backdrop of vigorous jet substructure explorations at the LHC, our systematic theoretical research will enable a deeper understanding of the recent ALICE results and provide more valuable references for future experimental programs.

\acknowledgments
This research is supported by the National Natural Science Foundation of China under Grant Nos.~12535010 and 12375137. S. W. is supported by the Open Foundation of Key Laboratory of Quark and Lepton Physics (MOE) No. QLPL2023P01 and the Talent Scientific Start-up Foundation of China Three Gorges University (CTGU) No. 2024RCKJ013.

\bigskip\noindent
\textbf{Conflict of Interest}\quad
The authors declare that they have no conflict of interest.

\appendix

\section{Medium modification of $k_{\rm T, g}$ distribution for photon-tagged jets}
\label{sec:gjet-ktg}

\begin{figure}[h]
\begin{center}
\includegraphics[width=2.3in,angle=0]{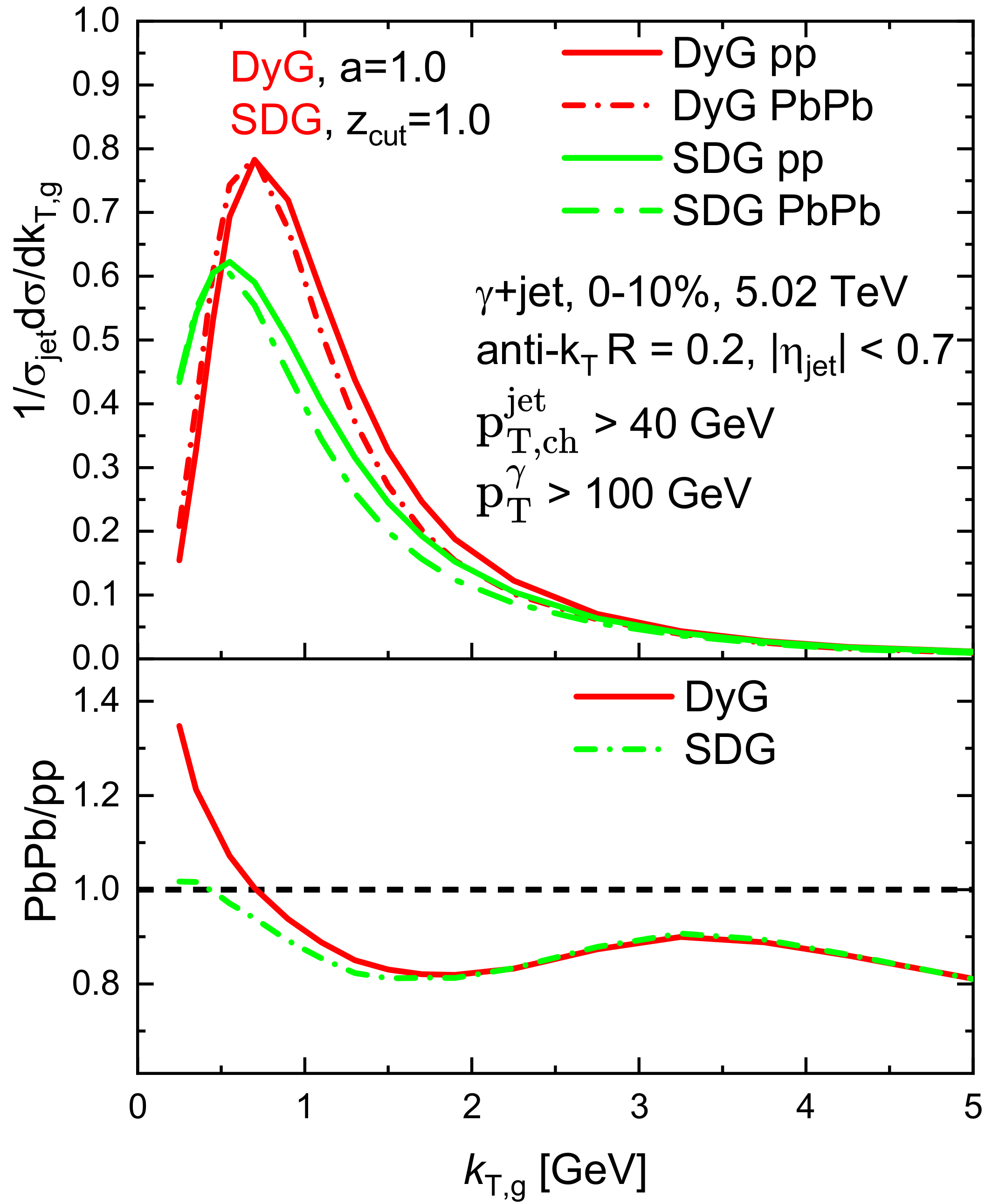}
\vspace*{-0in}
\caption{(Color online) $k_{\rm T, g}$ distributions of the groomed photon-tagged jets in pp and $0-10\%$ PbPb collisions at $\sqrt{s_{NN}}=5.02$ TeV utilizing both the Dynamical Grooming and Soft-Drop Grooming algorithms. The ratios of PbPb/pp are also shown in the bottom panels.}
\label{fig:gi-ktg}
\end{center}
\end{figure}

In the main text, the nuclear modification effect of $k_{T,g}$ for photon-tagged jets was not discussed, as it has been demonstrated that a broadened $k_{T,g}$ distribution cannot be observed in nucleus-nucleus collisions even with the complete elimination of selection bias. For the sake of rigor and completeness, we nonetheless present and discuss this result in the present appendix. As shown in Fig.~\ref{fig:gi-ktg}, the $k_{T,g}$ distributions for $\gamma+$jets in pp and $0-10\%$ central PbPb collisions at $\sqrt{s_{NN}}=5.02$ TeV, as well as their corresponding PbPb/pp ratios, are calculated using both the DyG and SDG algorithms. We find that, irrespective of the algorithm employed, the $k_{T,g}$ distribution in PbPb collisions is narrowed compared to that in pp collisions. This manifests as a significant suppression at large $k_{T,g}$ in the PbPb/pp ratio, a behavior similar to that observed for inclusive jets. Consequently, even with $\gamma+$jets, which substantially mitigate selection bias, observing a broadening in the $k_{T,g}$ distribution remains highly challenging. This constitutes the primary rationale for shifting our focus to the groomed radius $R_g$.

\bibliography{sarefs}

\begin{thebibliography}{110}%
\makeatletter
\providecommand \@ifxundefined [1]{%
 \@ifx{#1\undefined}
}%
\providecommand \@ifnum [1]{%
 \ifnum #1\expandafter \@firstoftwo
 \else \expandafter \@secondoftwo
 \fi
}%
\providecommand \@ifx [1]{%
 \ifx #1\expandafter \@firstoftwo
 \else \expandafter \@secondoftwo
 \fi
}%
\providecommand \natexlab [1]{#1}%
\providecommand \enquote  [1]{``#1''}%
\providecommand \bibnamefont  [1]{#1}%
\providecommand \bibfnamefont [1]{#1}%
\providecommand \citenamefont [1]{#1}%
\providecommand \href@noop [0]{\@secondoftwo}%
\providecommand \href [0]{\begingroup \@sanitize@url \@href}%
\providecommand \@href[1]{\@@startlink{#1}\@@href}%
\providecommand \@@href[1]{\endgroup#1\@@endlink}%
\providecommand \@sanitize@url [0]{\catcode `\\12\catcode `\$12\catcode
  `\&12\catcode `\#12\catcode `\^12\catcode `\_12\catcode `\%12\relax}%
\providecommand \@@startlink[1]{}%
\providecommand \@@endlink[0]{}%
\providecommand \url  [0]{\begingroup\@sanitize@url \@url }%
\providecommand \@url [1]{\endgroup\@href {#1}{\urlprefix }}%
\providecommand \urlprefix  [0]{URL }%
\providecommand \Eprint [0]{\href }%
\providecommand \doibase [0]{http://dx.doi.org/}%
\providecommand \selectlanguage [0]{\@gobble}%
\providecommand \bibinfo  [0]{\@secondoftwo}%
\providecommand \bibfield  [0]{\@secondoftwo}%
\providecommand \translation [1]{[#1]}%
\providecommand \BibitemOpen [0]{}%
\providecommand \bibitemStop [0]{}%
\providecommand \bibitemNoStop [0]{.\EOS\space}%
\providecommand \EOS [0]{\spacefactor3000\relax}%
\providecommand \BibitemShut  [1]{\csname bibitem#1\endcsname}%
\let\auto@bib@innerbib\@empty
\bibitem [{\citenamefont {Matsui}\ and\ \citenamefont
  {Satz}(1986)}]{Matsui:1986dk}%
  \BibitemOpen
  \bibfield  {author} {\bibinfo {author} {\bibfnamefont {T.}~\bibnamefont
  {Matsui}}\ and\ \bibinfo {author} {\bibfnamefont {H.}~\bibnamefont {Satz}},\
  }\href {\doibase 10.1016/0370-2693(86)91404-8} {\bibfield  {journal}
  {\bibinfo  {journal} {Phys. Lett. B}\ }\textbf {\bibinfo {volume} {178}},\
  \bibinfo {pages} {416} (\bibinfo {year} {1986})}\BibitemShut {NoStop}%
\bibitem [{\citenamefont {Wang}\ and\ \citenamefont
  {Gyulassy}(1991)}]{Wang:1991hta}%
  \BibitemOpen
  \bibfield  {author} {\bibinfo {author} {\bibfnamefont {X.-N.}\ \bibnamefont
  {Wang}}\ and\ \bibinfo {author} {\bibfnamefont {M.}~\bibnamefont
  {Gyulassy}},\ }\href {\doibase 10.1103/PhysRevD.44.3501} {\bibfield
  {journal} {\bibinfo  {journal} {Phys. Rev. D}\ }\textbf {\bibinfo {volume}
  {44}},\ \bibinfo {pages} {3501} (\bibinfo {year} {1991})}\BibitemShut
  {NoStop}%
\bibitem [{\citenamefont {Gyulassy}\ and\ \citenamefont
  {McLerran}(2005)}]{Gyulassy:2004zy}%
  \BibitemOpen
  \bibfield  {author} {\bibinfo {author} {\bibfnamefont {M.}~\bibnamefont
  {Gyulassy}}\ and\ \bibinfo {author} {\bibfnamefont {L.}~\bibnamefont
  {McLerran}},\ }\href {\doibase 10.1016/j.nuclphysa.2004.10.034} {\bibfield
  {journal} {\bibinfo  {journal} {Nucl. Phys. A}\ }\textbf {\bibinfo {volume}
  {750}},\ \bibinfo {pages} {30} (\bibinfo {year} {2005})},\ \Eprint
  {http://arxiv.org/abs/nucl-th/0405013} {arXiv:nucl-th/0405013} \BibitemShut
  {NoStop}%
\bibitem [{\citenamefont {Adcox}\ \emph {et~al.}(2005)\citenamefont {Adcox}
  \emph {et~al.}}]{PHENIX:2004vcz}%
  \BibitemOpen
  \bibfield  {author} {\bibinfo {author} {\bibfnamefont {K.}~\bibnamefont
  {Adcox}} \emph {et~al.} (\bibinfo {collaboration} {PHENIX}),\ }\href
  {\doibase 10.1016/j.nuclphysa.2005.03.086} {\bibfield  {journal} {\bibinfo
  {journal} {Nucl. Phys. A}\ }\textbf {\bibinfo {volume} {757}},\ \bibinfo
  {pages} {184} (\bibinfo {year} {2005})},\ \Eprint
  {http://arxiv.org/abs/nucl-ex/0410003} {arXiv:nucl-ex/0410003} \BibitemShut
  {NoStop}%
\bibitem [{\citenamefont {Adams}\ \emph {et~al.}(2005)\citenamefont {Adams}
  \emph {et~al.}}]{STAR:2005gfr}%
  \BibitemOpen
  \bibfield  {author} {\bibinfo {author} {\bibfnamefont {J.}~\bibnamefont
  {Adams}} \emph {et~al.} (\bibinfo {collaboration} {STAR}),\ }\href {\doibase
  10.1016/j.nuclphysa.2005.03.085} {\bibfield  {journal} {\bibinfo  {journal}
  {Nucl. Phys. A}\ }\textbf {\bibinfo {volume} {757}},\ \bibinfo {pages} {102}
  (\bibinfo {year} {2005})},\ \Eprint {http://arxiv.org/abs/nucl-ex/0501009}
  {arXiv:nucl-ex/0501009} \BibitemShut {NoStop}%
\bibitem [{\citenamefont {Chen}\ \emph {et~al.}(2024)\citenamefont {Chen} \emph
  {et~al.}}]{Chen:2024aom}%
  \BibitemOpen
  \bibfield  {author} {\bibinfo {author} {\bibfnamefont {J.}~\bibnamefont
  {Chen}} \emph {et~al.},\ }\href {\doibase 10.1007/s41365-024-01591-2}
  {\bibfield  {journal} {\bibinfo  {journal} {Nucl. Sci. Tech.}\ }\textbf
  {\bibinfo {volume} {35}},\ \bibinfo {pages} {214} (\bibinfo {year} {2024})},\
  \Eprint {http://arxiv.org/abs/2407.02935} {arXiv:2407.02935 [nucl-ex]}
  \BibitemShut {NoStop}%
\bibitem [{\citenamefont {Shou}\ \emph {et~al.}(2024)\citenamefont {Shou} \emph
  {et~al.}}]{Shou:2024uga}%
  \BibitemOpen
  \bibfield  {author} {\bibinfo {author} {\bibfnamefont {Q.-Y.}\ \bibnamefont
  {Shou}} \emph {et~al.},\ }\href {\doibase 10.1007/s41365-024-01583-2}
  {\bibfield  {journal} {\bibinfo  {journal} {Nucl. Sci. Tech.}\ }\textbf
  {\bibinfo {volume} {35}},\ \bibinfo {pages} {219} (\bibinfo {year} {2024})},\
  \Eprint {http://arxiv.org/abs/2409.17964} {arXiv:2409.17964 [nucl-ex]}
  \BibitemShut {NoStop}%
\bibitem [{\citenamefont {Wu}(2025)}]{Wu:2025xpz}%
  \BibitemOpen
  \bibfield  {author} {\bibinfo {author} {\bibfnamefont {S.}~\bibnamefont
  {Wu}},\ }\href {\doibase 10.1007/s41365-025-01721-4} {\bibfield  {journal}
  {\bibinfo  {journal} {Nucl. Sci. Tech.}\ }\textbf {\bibinfo {volume} {36}},\
  \bibinfo {pages} {149} (\bibinfo {year} {2025})}\BibitemShut {NoStop}%
\bibitem [{\citenamefont {Zhu}\ \emph {et~al.}(2026)\citenamefont {Zhu},
  \citenamefont {Luo}, \citenamefont {Chen}, \citenamefont {Zhou},
  \citenamefont {Zhang},\ and\ \citenamefont {Hou}}]{Zhu:2025xiz}%
  \BibitemOpen
  \bibfield  {author} {\bibinfo {author} {\bibfnamefont {L.-Q.}\ \bibnamefont
  {Zhu}}, \bibinfo {author} {\bibfnamefont {O.-Y.}\ \bibnamefont {Luo}},
  \bibinfo {author} {\bibfnamefont {X.}~\bibnamefont {Chen}}, \bibinfo {author}
  {\bibfnamefont {K.}~\bibnamefont {Zhou}}, \bibinfo {author} {\bibfnamefont
  {H.-Z.}\ \bibnamefont {Zhang}}, \ and\ \bibinfo {author} {\bibfnamefont
  {D.-F.}\ \bibnamefont {Hou}},\ }\href {\doibase 10.1007/s41365-026-01903-8}
  {\bibfield  {journal} {\bibinfo  {journal} {Nucl. Sci. Tech.}\ }\textbf
  {\bibinfo {volume} {37}},\ \bibinfo {pages} {68} (\bibinfo {year} {2026})},\
  \Eprint {http://arxiv.org/abs/2508.12756} {arXiv:2508.12756 [hep-ph]}
  \BibitemShut {NoStop}%
\bibitem [{\citenamefont {Ma}(2026)}]{Ma:2026xiz}%
  \BibitemOpen
  \bibfield  {author} {\bibinfo {author} {\bibfnamefont {Y.-G.}\ \bibnamefont
  {Ma}},\ }\href {\doibase 10.1007/s41365-025-01859-1} {\bibfield  {journal}
  {\bibinfo  {journal} {Nucl. Sci. Tech.}\ }\textbf {\bibinfo {volume} {37}},\
  \bibinfo {pages} {51} (\bibinfo {year} {2026})}\BibitemShut {NoStop}%
\bibitem [{\citenamefont {Braaten}\ and\ \citenamefont
  {Thoma}(1991)}]{Braaten:1991we}%
  \BibitemOpen
  \bibfield  {author} {\bibinfo {author} {\bibfnamefont {E.}~\bibnamefont
  {Braaten}}\ and\ \bibinfo {author} {\bibfnamefont {M.~H.}\ \bibnamefont
  {Thoma}},\ }\href {\doibase 10.1103/PhysRevD.44.R2625} {\bibfield  {journal}
  {\bibinfo  {journal} {Phys. Rev. D}\ }\textbf {\bibinfo {volume} {44}},\
  \bibinfo {pages} {R2625} (\bibinfo {year} {1991})}\BibitemShut {NoStop}%
\bibitem [{\citenamefont {Baier}\ \emph
  {et~al.}(1997{\natexlab{a}})\citenamefont {Baier}, \citenamefont
  {Dokshitzer}, \citenamefont {Mueller}, \citenamefont {Peigne},\ and\
  \citenamefont {Schiff}}]{Baier:1996kr}%
  \BibitemOpen
  \bibfield  {author} {\bibinfo {author} {\bibfnamefont {R.}~\bibnamefont
  {Baier}}, \bibinfo {author} {\bibfnamefont {Y.~L.}\ \bibnamefont
  {Dokshitzer}}, \bibinfo {author} {\bibfnamefont {A.~H.}\ \bibnamefont
  {Mueller}}, \bibinfo {author} {\bibfnamefont {S.}~\bibnamefont {Peigne}}, \
  and\ \bibinfo {author} {\bibfnamefont {D.}~\bibnamefont {Schiff}},\ }\href
  {\doibase 10.1016/S0550-3213(96)00553-6} {\bibfield  {journal} {\bibinfo
  {journal} {Nucl. Phys. B}\ }\textbf {\bibinfo {volume} {483}},\ \bibinfo
  {pages} {291} (\bibinfo {year} {1997}{\natexlab{a}})},\ \Eprint
  {http://arxiv.org/abs/hep-ph/9607355} {arXiv:hep-ph/9607355} \BibitemShut
  {NoStop}%
\bibitem [{\citenamefont {Baier}\ \emph
  {et~al.}(1997{\natexlab{b}})\citenamefont {Baier}, \citenamefont
  {Dokshitzer}, \citenamefont {Mueller}, \citenamefont {Peigne},\ and\
  \citenamefont {Schiff}}]{Baier:1996sk}%
  \BibitemOpen
  \bibfield  {author} {\bibinfo {author} {\bibfnamefont {R.}~\bibnamefont
  {Baier}}, \bibinfo {author} {\bibfnamefont {Y.~L.}\ \bibnamefont
  {Dokshitzer}}, \bibinfo {author} {\bibfnamefont {A.~H.}\ \bibnamefont
  {Mueller}}, \bibinfo {author} {\bibfnamefont {S.}~\bibnamefont {Peigne}}, \
  and\ \bibinfo {author} {\bibfnamefont {D.}~\bibnamefont {Schiff}},\ }\href
  {\doibase 10.1016/S0550-3213(96)00581-0} {\bibfield  {journal} {\bibinfo
  {journal} {Nucl. Phys. B}\ }\textbf {\bibinfo {volume} {484}},\ \bibinfo
  {pages} {265} (\bibinfo {year} {1997}{\natexlab{b}})},\ \Eprint
  {http://arxiv.org/abs/hep-ph/9608322} {arXiv:hep-ph/9608322} \BibitemShut
  {NoStop}%
\bibitem [{\citenamefont {Zakharov}(1996)}]{Zakharov:1996fv}%
  \BibitemOpen
  \bibfield  {author} {\bibinfo {author} {\bibfnamefont {B.~G.}\ \bibnamefont
  {Zakharov}},\ }\href {\doibase 10.1134/1.567126} {\bibfield  {journal}
  {\bibinfo  {journal} {JETP Lett.}\ }\textbf {\bibinfo {volume} {63}},\
  \bibinfo {pages} {952} (\bibinfo {year} {1996})},\ \Eprint
  {http://arxiv.org/abs/hep-ph/9607440} {arXiv:hep-ph/9607440} \BibitemShut
  {NoStop}%
\bibitem [{\citenamefont {Gyulassy}\ \emph {et~al.}(2000)\citenamefont
  {Gyulassy}, \citenamefont {Levai},\ and\ \citenamefont
  {Vitev}}]{Gyulassy:1999zd}%
  \BibitemOpen
  \bibfield  {author} {\bibinfo {author} {\bibfnamefont {M.}~\bibnamefont
  {Gyulassy}}, \bibinfo {author} {\bibfnamefont {P.}~\bibnamefont {Levai}}, \
  and\ \bibinfo {author} {\bibfnamefont {I.}~\bibnamefont {Vitev}},\ }\href
  {\doibase 10.1016/S0550-3213(99)00713-0} {\bibfield  {journal} {\bibinfo
  {journal} {Nucl. Phys. B}\ }\textbf {\bibinfo {volume} {571}},\ \bibinfo
  {pages} {197} (\bibinfo {year} {2000})},\ \Eprint
  {http://arxiv.org/abs/hep-ph/9907461} {arXiv:hep-ph/9907461} \BibitemShut
  {NoStop}%
\bibitem [{\citenamefont {Wiedemann}(2000)}]{Wiedemann:2000za}%
  \BibitemOpen
  \bibfield  {author} {\bibinfo {author} {\bibfnamefont {U.~A.}\ \bibnamefont
  {Wiedemann}},\ }\href {\doibase 10.1016/S0550-3213(00)00457-0} {\bibfield
  {journal} {\bibinfo  {journal} {Nucl. Phys. B}\ }\textbf {\bibinfo {volume}
  {588}},\ \bibinfo {pages} {303} (\bibinfo {year} {2000})},\ \Eprint
  {http://arxiv.org/abs/hep-ph/0005129} {arXiv:hep-ph/0005129} \BibitemShut
  {NoStop}%
\bibitem [{\citenamefont {Arnold}\ \emph
  {et~al.}(2001{\natexlab{a}})\citenamefont {Arnold}, \citenamefont {Moore},\
  and\ \citenamefont {Yaffe}}]{Arnold:2001ba}%
  \BibitemOpen
  \bibfield  {author} {\bibinfo {author} {\bibfnamefont {P.~B.}\ \bibnamefont
  {Arnold}}, \bibinfo {author} {\bibfnamefont {G.~D.}\ \bibnamefont {Moore}}, \
  and\ \bibinfo {author} {\bibfnamefont {L.~G.}\ \bibnamefont {Yaffe}},\ }\href
  {\doibase 10.1088/1126-6708/2001/11/057} {\bibfield  {journal} {\bibinfo
  {journal} {JHEP}\ }\textbf {\bibinfo {volume} {11}},\ \bibinfo {pages} {057}
  (\bibinfo {year} {2001}{\natexlab{a}})},\ \Eprint
  {http://arxiv.org/abs/hep-ph/0109064} {arXiv:hep-ph/0109064} \BibitemShut
  {NoStop}%
\bibitem [{\citenamefont {Arnold}\ \emph
  {et~al.}(2001{\natexlab{b}})\citenamefont {Arnold}, \citenamefont {Moore},\
  and\ \citenamefont {Yaffe}}]{Arnold:2001ms}%
  \BibitemOpen
  \bibfield  {author} {\bibinfo {author} {\bibfnamefont {P.~B.}\ \bibnamefont
  {Arnold}}, \bibinfo {author} {\bibfnamefont {G.~D.}\ \bibnamefont {Moore}}, \
  and\ \bibinfo {author} {\bibfnamefont {L.~G.}\ \bibnamefont {Yaffe}},\ }\href
  {\doibase 10.1088/1126-6708/2001/12/009} {\bibfield  {journal} {\bibinfo
  {journal} {JHEP}\ }\textbf {\bibinfo {volume} {12}},\ \bibinfo {pages} {009}
  (\bibinfo {year} {2001}{\natexlab{b}})},\ \Eprint
  {http://arxiv.org/abs/hep-ph/0111107} {arXiv:hep-ph/0111107} \BibitemShut
  {NoStop}%
\bibitem [{\citenamefont {Wang}\ and\ \citenamefont
  {Guo}(2001)}]{Wang:2001ifa}%
  \BibitemOpen
  \bibfield  {author} {\bibinfo {author} {\bibfnamefont {X.-N.}\ \bibnamefont
  {Wang}}\ and\ \bibinfo {author} {\bibfnamefont {X.-f.}\ \bibnamefont {Guo}},\
  }\href {\doibase 10.1016/S0375-9474(01)01130-7} {\bibfield  {journal}
  {\bibinfo  {journal} {Nucl. Phys. A}\ }\textbf {\bibinfo {volume} {696}},\
  \bibinfo {pages} {788} (\bibinfo {year} {2001})},\ \Eprint
  {http://arxiv.org/abs/hep-ph/0102230} {arXiv:hep-ph/0102230} \BibitemShut
  {NoStop}%
\bibitem [{\citenamefont {Vitev}\ and\ \citenamefont
  {Zhang}(2010)}]{Vitev:2009rd}%
  \BibitemOpen
  \bibfield  {author} {\bibinfo {author} {\bibfnamefont {I.}~\bibnamefont
  {Vitev}}\ and\ \bibinfo {author} {\bibfnamefont {B.-W.}\ \bibnamefont
  {Zhang}},\ }\href {\doibase 10.1103/PhysRevLett.104.132001} {\bibfield
  {journal} {\bibinfo  {journal} {Phys. Rev. Lett.}\ }\textbf {\bibinfo
  {volume} {104}},\ \bibinfo {pages} {132001} (\bibinfo {year} {2010})},\
  \Eprint {http://arxiv.org/abs/0910.1090} {arXiv:0910.1090 [hep-ph]}
  \BibitemShut {NoStop}%
\bibitem [{\citenamefont {Gyulassy}\ and\ \citenamefont
  {Wang}(1994)}]{Gyulassy:1993hr}%
  \BibitemOpen
  \bibfield  {author} {\bibinfo {author} {\bibfnamefont {M.}~\bibnamefont
  {Gyulassy}}\ and\ \bibinfo {author} {\bibfnamefont {X.-n.}\ \bibnamefont
  {Wang}},\ }\href {\doibase 10.1016/0550-3213(94)90079-5} {\bibfield
  {journal} {\bibinfo  {journal} {Nucl. Phys. B}\ }\textbf {\bibinfo {volume}
  {420}},\ \bibinfo {pages} {583} (\bibinfo {year} {1994})},\ \Eprint
  {http://arxiv.org/abs/nucl-th/9306003} {arXiv:nucl-th/9306003} \BibitemShut
  {NoStop}%
\bibitem [{\citenamefont {Gyulassy}\ \emph {et~al.}(2004)\citenamefont
  {Gyulassy}, \citenamefont {Vitev}, \citenamefont {Wang},\ and\ \citenamefont
  {Zhang}}]{Gyulassy:2003mc}%
  \BibitemOpen
  \bibfield  {author} {\bibinfo {author} {\bibfnamefont {M.}~\bibnamefont
  {Gyulassy}}, \bibinfo {author} {\bibfnamefont {I.}~\bibnamefont {Vitev}},
  \bibinfo {author} {\bibfnamefont {X.-N.}\ \bibnamefont {Wang}}, \ and\
  \bibinfo {author} {\bibfnamefont {B.-W.}\ \bibnamefont {Zhang}},\ }\href
  {\doibase 10.1142/9789812795533_0003} {\ ,\ \bibinfo {pages} {123} (\bibinfo
  {year} {2004})},\ \Eprint {http://arxiv.org/abs/nucl-th/0302077}
  {arXiv:nucl-th/0302077} \BibitemShut {NoStop}%
\bibitem [{\citenamefont {Qin}\ and\ \citenamefont {Wang}(2015)}]{Qin:2015srf}%
  \BibitemOpen
  \bibfield  {author} {\bibinfo {author} {\bibfnamefont {G.-Y.}\ \bibnamefont
  {Qin}}\ and\ \bibinfo {author} {\bibfnamefont {X.-N.}\ \bibnamefont {Wang}},\
  }\href {\doibase 10.1142/S0218301315300143} {\bibfield  {journal} {\bibinfo
  {journal} {Int. J. Mod. Phys. E}\ }\textbf {\bibinfo {volume} {24}},\
  \bibinfo {pages} {1530014} (\bibinfo {year} {2015})},\ \Eprint
  {http://arxiv.org/abs/1511.00790} {arXiv:1511.00790 [hep-ph]} \BibitemShut
  {NoStop}%
\bibitem [{\citenamefont {Xie}\ \emph {et~al.}(2024)\citenamefont {Xie},
  \citenamefont {Han}, \citenamefont {Wang}, \citenamefont {Zhang},\ and\
  \citenamefont {Zhang}}]{Xie:2024xbn}%
  \BibitemOpen
  \bibfield  {author} {\bibinfo {author} {\bibfnamefont {M.}~\bibnamefont
  {Xie}}, \bibinfo {author} {\bibfnamefont {Q.-F.}\ \bibnamefont {Han}},
  \bibinfo {author} {\bibfnamefont {E.-K.}\ \bibnamefont {Wang}}, \bibinfo
  {author} {\bibfnamefont {B.-W.}\ \bibnamefont {Zhang}}, \ and\ \bibinfo
  {author} {\bibfnamefont {H.-Z.}\ \bibnamefont {Zhang}},\ }\href {\doibase
  10.1007/s41365-024-01492-4} {\bibfield  {journal} {\bibinfo  {journal} {Nucl.
  Sci. Tech.}\ }\textbf {\bibinfo {volume} {35}},\ \bibinfo {pages} {125}
  (\bibinfo {year} {2024})},\ \Eprint {http://arxiv.org/abs/2409.18773}
  {arXiv:2409.18773 [hep-ph]} \BibitemShut {NoStop}%
\bibitem [{\citenamefont {Yang}\ \emph {et~al.}(2025)\citenamefont {Yang},
  \citenamefont {Kong}, \citenamefont {Ru},\ and\ \citenamefont
  {Zhang}}]{Yang:2025yaf}%
  \BibitemOpen
  \bibfield  {author} {\bibinfo {author} {\bibfnamefont {M.-Q.}\ \bibnamefont
  {Yang}}, \bibinfo {author} {\bibfnamefont {W.-X.}\ \bibnamefont {Kong}},
  \bibinfo {author} {\bibfnamefont {P.}~\bibnamefont {Ru}}, \ and\ \bibinfo
  {author} {\bibfnamefont {B.-W.}\ \bibnamefont {Zhang}},\ }\href {\doibase
  10.1016/j.physletb.2025.139746} {\bibfield  {journal} {\bibinfo  {journal}
  {Phys. Lett. B}\ }\textbf {\bibinfo {volume} {868}},\ \bibinfo {pages}
  {139746} (\bibinfo {year} {2025})},\ \Eprint
  {http://arxiv.org/abs/2503.15042} {arXiv:2503.15042 [hep-ph]} \BibitemShut
  {NoStop}%
\bibitem [{\citenamefont {Kong}\ and\ \citenamefont
  {Zhang}(2025)}]{Kong:2024wdk}%
  \BibitemOpen
  \bibfield  {author} {\bibinfo {author} {\bibfnamefont {W.-X.}\ \bibnamefont
  {Kong}}\ and\ \bibinfo {author} {\bibfnamefont {B.-W.}\ \bibnamefont
  {Zhang}},\ }\href {\doibase 10.1088/1674-1137/add5d8} {\bibfield  {journal}
  {\bibinfo  {journal} {Chin. Phys.}\ }\textbf {\bibinfo {volume} {49}},\
  \bibinfo {pages} {094101} (\bibinfo {year} {2025})},\ \Eprint
  {http://arxiv.org/abs/2407.20680} {arXiv:2407.20680 [hep-ph]} \BibitemShut
  {NoStop}%
\bibitem [{\citenamefont {Kang}\ \emph
  {et~al.}(2025{\natexlab{a}})\citenamefont {Kang}, \citenamefont {Wang},
  \citenamefont {Dai}, \citenamefont {Wang},\ and\ \citenamefont
  {Zhang}}]{Kang:2023qxb}%
  \BibitemOpen
  \bibfield  {author} {\bibinfo {author} {\bibfnamefont {J.-W.}\ \bibnamefont
  {Kang}}, \bibinfo {author} {\bibfnamefont {L.}~\bibnamefont {Wang}}, \bibinfo
  {author} {\bibfnamefont {W.}~\bibnamefont {Dai}}, \bibinfo {author}
  {\bibfnamefont {S.}~\bibnamefont {Wang}}, \ and\ \bibinfo {author}
  {\bibfnamefont {B.-W.}\ \bibnamefont {Zhang}},\ }\href {\doibase
  10.1103/71sd-7qqb} {\bibfield  {journal} {\bibinfo  {journal} {Phys. Rev. C}\
  }\textbf {\bibinfo {volume} {112}},\ \bibinfo {pages} {034903} (\bibinfo
  {year} {2025}{\natexlab{a}})},\ \Eprint {http://arxiv.org/abs/2304.04649}
  {arXiv:2304.04649 [nucl-th]} \BibitemShut {NoStop}%
\bibitem [{\citenamefont {He}\ and\ \citenamefont {Rapp}(2023)}]{He:2022tod}%
  \BibitemOpen
  \bibfield  {author} {\bibinfo {author} {\bibfnamefont {M.}~\bibnamefont
  {He}}\ and\ \bibinfo {author} {\bibfnamefont {R.}~\bibnamefont {Rapp}},\
  }\href {\doibase 10.1103/PhysRevLett.131.012301} {\bibfield  {journal}
  {\bibinfo  {journal} {Phys. Rev. Lett.}\ }\textbf {\bibinfo {volume} {131}},\
  \bibinfo {pages} {012301} (\bibinfo {year} {2023})},\ \Eprint
  {http://arxiv.org/abs/2209.13419} {arXiv:2209.13419 [hep-ph]} \BibitemShut
  {NoStop}%
\bibitem [{\citenamefont {Barata}\ \emph {et~al.}(2022)\citenamefont {Barata},
  \citenamefont {Sadofyev},\ and\ \citenamefont {Salgado}}]{Barata:2022krd}%
  \BibitemOpen
  \bibfield  {author} {\bibinfo {author} {\bibfnamefont {J.}~\bibnamefont
  {Barata}}, \bibinfo {author} {\bibfnamefont {A.~V.}\ \bibnamefont
  {Sadofyev}}, \ and\ \bibinfo {author} {\bibfnamefont {C.~A.}\ \bibnamefont
  {Salgado}},\ }\href {\doibase 10.1103/PhysRevD.105.114010} {\bibfield
  {journal} {\bibinfo  {journal} {Phys. Rev. D}\ }\textbf {\bibinfo {volume}
  {105}},\ \bibinfo {pages} {114010} (\bibinfo {year} {2022})},\ \Eprint
  {http://arxiv.org/abs/2202.08847} {arXiv:2202.08847 [hep-ph]} \BibitemShut
  {NoStop}%
\bibitem [{\citenamefont {Andres}\ \emph {et~al.}(2022)\citenamefont {Andres},
  \citenamefont {Dominguez}, \citenamefont {Sadofyev},\ and\ \citenamefont
  {Salgado}}]{Andres:2022ndd}%
  \BibitemOpen
  \bibfield  {author} {\bibinfo {author} {\bibfnamefont {C.}~\bibnamefont
  {Andres}}, \bibinfo {author} {\bibfnamefont {F.}~\bibnamefont {Dominguez}},
  \bibinfo {author} {\bibfnamefont {A.~V.}\ \bibnamefont {Sadofyev}}, \ and\
  \bibinfo {author} {\bibfnamefont {C.~A.}\ \bibnamefont {Salgado}},\ }\href
  {\doibase 10.1103/PhysRevD.106.074023} {\bibfield  {journal} {\bibinfo
  {journal} {Phys. Rev. D}\ }\textbf {\bibinfo {volume} {106}},\ \bibinfo
  {pages} {074023} (\bibinfo {year} {2022})},\ \Eprint
  {http://arxiv.org/abs/2207.07141} {arXiv:2207.07141 [hep-ph]} \BibitemShut
  {NoStop}%
\bibitem [{\citenamefont {Kunnawalkam~Elayavalli}\ and\ \citenamefont
  {Zapp}(2017)}]{KunnawalkamElayavalli:2017hxo}%
  \BibitemOpen
  \bibfield  {author} {\bibinfo {author} {\bibfnamefont {R.}~\bibnamefont
  {Kunnawalkam~Elayavalli}}\ and\ \bibinfo {author} {\bibfnamefont {K.~C.}\
  \bibnamefont {Zapp}},\ }\href {\doibase 10.1007/JHEP07(2017)141} {\bibfield
  {journal} {\bibinfo  {journal} {JHEP}\ }\textbf {\bibinfo {volume} {07}},\
  \bibinfo {pages} {141} (\bibinfo {year} {2017})},\ \Eprint
  {http://arxiv.org/abs/1707.01539} {arXiv:1707.01539 [hep-ph]} \BibitemShut
  {NoStop}%
\bibitem [{\citenamefont {Pablos}(2020)}]{Pablos:2019ngg}%
  \BibitemOpen
  \bibfield  {author} {\bibinfo {author} {\bibfnamefont {D.}~\bibnamefont
  {Pablos}},\ }\href {\doibase 10.1103/PhysRevLett.124.052301} {\bibfield
  {journal} {\bibinfo  {journal} {Phys. Rev. Lett.}\ }\textbf {\bibinfo
  {volume} {124}},\ \bibinfo {pages} {052301} (\bibinfo {year} {2020})},\
  \Eprint {http://arxiv.org/abs/1907.12301} {arXiv:1907.12301 [hep-ph]}
  \BibitemShut {NoStop}%
\bibitem [{\citenamefont {Chen}\ \emph {et~al.}(2020)\citenamefont {Chen},
  \citenamefont {Cao}, \citenamefont {Luo}, \citenamefont {Pang},\ and\
  \citenamefont {Wang}}]{Chen:2020tbl}%
  \BibitemOpen
  \bibfield  {author} {\bibinfo {author} {\bibfnamefont {W.}~\bibnamefont
  {Chen}}, \bibinfo {author} {\bibfnamefont {S.}~\bibnamefont {Cao}}, \bibinfo
  {author} {\bibfnamefont {T.}~\bibnamefont {Luo}}, \bibinfo {author}
  {\bibfnamefont {L.-G.}\ \bibnamefont {Pang}}, \ and\ \bibinfo {author}
  {\bibfnamefont {X.-N.}\ \bibnamefont {Wang}},\ }\href {\doibase
  10.1016/j.physletb.2020.135783} {\bibfield  {journal} {\bibinfo  {journal}
  {Phys. Lett. B}\ }\textbf {\bibinfo {volume} {810}},\ \bibinfo {pages}
  {135783} (\bibinfo {year} {2020})},\ \Eprint
  {http://arxiv.org/abs/2005.09678} {arXiv:2005.09678 [hep-ph]} \BibitemShut
  {NoStop}%
\bibitem [{\citenamefont {Casalderrey-Solana}\ \emph
  {et~al.}(2021)\citenamefont {Casalderrey-Solana}, \citenamefont {Milhano},
  \citenamefont {Pablos}, \citenamefont {Rajagopal},\ and\ \citenamefont
  {Yao}}]{Casalderrey-Solana:2020rsj}%
  \BibitemOpen
  \bibfield  {author} {\bibinfo {author} {\bibfnamefont {J.}~\bibnamefont
  {Casalderrey-Solana}}, \bibinfo {author} {\bibfnamefont {J.~G.}\ \bibnamefont
  {Milhano}}, \bibinfo {author} {\bibfnamefont {D.}~\bibnamefont {Pablos}},
  \bibinfo {author} {\bibfnamefont {K.}~\bibnamefont {Rajagopal}}, \ and\
  \bibinfo {author} {\bibfnamefont {X.}~\bibnamefont {Yao}},\ }\href {\doibase
  10.1007/JHEP05(2021)230} {\bibfield  {journal} {\bibinfo  {journal} {JHEP}\
  }\textbf {\bibinfo {volume} {05}},\ \bibinfo {pages} {230} (\bibinfo {year}
  {2021})},\ \Eprint {http://arxiv.org/abs/2010.01140} {arXiv:2010.01140
  [hep-ph]} \BibitemShut {NoStop}%
\bibitem [{\citenamefont {He}\ \emph {et~al.}(2019)\citenamefont {He},
  \citenamefont {Cao}, \citenamefont {Chen}, \citenamefont {Luo}, \citenamefont
  {Pang},\ and\ \citenamefont {Wang}}]{He:2018xjv}%
  \BibitemOpen
  \bibfield  {author} {\bibinfo {author} {\bibfnamefont {Y.}~\bibnamefont
  {He}}, \bibinfo {author} {\bibfnamefont {S.}~\bibnamefont {Cao}}, \bibinfo
  {author} {\bibfnamefont {W.}~\bibnamefont {Chen}}, \bibinfo {author}
  {\bibfnamefont {T.}~\bibnamefont {Luo}}, \bibinfo {author} {\bibfnamefont
  {L.-G.}\ \bibnamefont {Pang}}, \ and\ \bibinfo {author} {\bibfnamefont
  {X.-N.}\ \bibnamefont {Wang}},\ }\href {\doibase 10.1103/PhysRevC.99.054911}
  {\bibfield  {journal} {\bibinfo  {journal} {Phys. Rev. C}\ }\textbf {\bibinfo
  {volume} {99}},\ \bibinfo {pages} {054911} (\bibinfo {year} {2019})},\
  \Eprint {http://arxiv.org/abs/1809.02525} {arXiv:1809.02525 [nucl-th]}
  \BibitemShut {NoStop}%
\bibitem [{\citenamefont {Ke}\ and\ \citenamefont {Wang}(2021)}]{Ke:2020clc}%
  \BibitemOpen
  \bibfield  {author} {\bibinfo {author} {\bibfnamefont {W.}~\bibnamefont
  {Ke}}\ and\ \bibinfo {author} {\bibfnamefont {X.-N.}\ \bibnamefont {Wang}},\
  }\href {\doibase 10.1007/JHEP05(2021)041} {\bibfield  {journal} {\bibinfo
  {journal} {JHEP}\ }\textbf {\bibinfo {volume} {05}},\ \bibinfo {pages} {041}
  (\bibinfo {year} {2021})},\ \Eprint {http://arxiv.org/abs/2010.13680}
  {arXiv:2010.13680 [hep-ph]} \BibitemShut {NoStop}%
\bibitem [{\citenamefont {D'Eramo}\ \emph {et~al.}(2013)\citenamefont
  {D'Eramo}, \citenamefont {Lekaveckas}, \citenamefont {Liu},\ and\
  \citenamefont {Rajagopal}}]{DEramo:2012uzl}%
  \BibitemOpen
  \bibfield  {author} {\bibinfo {author} {\bibfnamefont {F.}~\bibnamefont
  {D'Eramo}}, \bibinfo {author} {\bibfnamefont {M.}~\bibnamefont {Lekaveckas}},
  \bibinfo {author} {\bibfnamefont {H.}~\bibnamefont {Liu}}, \ and\ \bibinfo
  {author} {\bibfnamefont {K.}~\bibnamefont {Rajagopal}},\ }\href {\doibase
  10.1007/JHEP05(2013)031} {\bibfield  {journal} {\bibinfo  {journal} {JHEP}\
  }\textbf {\bibinfo {volume} {05}},\ \bibinfo {pages} {031} (\bibinfo {year}
  {2013})},\ \Eprint {http://arxiv.org/abs/1211.1922} {arXiv:1211.1922
  [hep-ph]} \BibitemShut {NoStop}%
\bibitem [{\citenamefont {D'Eramo}\ \emph {et~al.}(2019)\citenamefont
  {D'Eramo}, \citenamefont {Rajagopal},\ and\ \citenamefont
  {Yin}}]{DEramo:2018eoy}%
  \BibitemOpen
  \bibfield  {author} {\bibinfo {author} {\bibfnamefont {F.}~\bibnamefont
  {D'Eramo}}, \bibinfo {author} {\bibfnamefont {K.}~\bibnamefont {Rajagopal}},
  \ and\ \bibinfo {author} {\bibfnamefont {Y.}~\bibnamefont {Yin}},\ }\href
  {\doibase 10.1007/JHEP01(2019)172} {\bibfield  {journal} {\bibinfo  {journal}
  {JHEP}\ }\textbf {\bibinfo {volume} {01}},\ \bibinfo {pages} {172} (\bibinfo
  {year} {2019})},\ \Eprint {http://arxiv.org/abs/1808.03250} {arXiv:1808.03250
  [hep-ph]} \BibitemShut {NoStop}%
\bibitem [{\citenamefont {Marzani}\ \emph {et~al.}(2019)\citenamefont
  {Marzani}, \citenamefont {Soyez},\ and\ \citenamefont
  {Spannowsky}}]{Marzani:2019hun}%
  \BibitemOpen
  \bibfield  {author} {\bibinfo {author} {\bibfnamefont {S.}~\bibnamefont
  {Marzani}}, \bibinfo {author} {\bibfnamefont {G.}~\bibnamefont {Soyez}}, \
  and\ \bibinfo {author} {\bibfnamefont {M.}~\bibnamefont {Spannowsky}},\
  }\href {\doibase 10.1007/978-3-030-15709-8} {\emph {\bibinfo {title}
  {{Looking inside jets: an introduction to jet substructure and boosted-object
  phenomenology}}}},\ Vol.\ \bibinfo {volume} {958}\ (\bibinfo  {publisher}
  {Springer},\ \bibinfo {year} {2019})\ \Eprint
  {http://arxiv.org/abs/1901.10342} {arXiv:1901.10342 [hep-ph]} \BibitemShut
  {NoStop}%
\bibitem [{\citenamefont {Cunqueiro}\ and\ \citenamefont
  {Sickles}(2022)}]{Cunqueiro:2021wls}%
  \BibitemOpen
  \bibfield  {author} {\bibinfo {author} {\bibfnamefont {L.}~\bibnamefont
  {Cunqueiro}}\ and\ \bibinfo {author} {\bibfnamefont {A.~M.}\ \bibnamefont
  {Sickles}},\ }\href {\doibase 10.1016/j.ppnp.2022.103940} {\bibfield
  {journal} {\bibinfo  {journal} {Prog. Part. Nucl. Phys.}\ }\textbf {\bibinfo
  {volume} {124}},\ \bibinfo {pages} {103940} (\bibinfo {year} {2022})},\
  \Eprint {http://arxiv.org/abs/2110.14490} {arXiv:2110.14490 [nucl-ex]}
  \BibitemShut {NoStop}%
\bibitem [{\citenamefont {Apolin{\'a}rio}\ \emph {et~al.}(2022)\citenamefont
  {Apolin{\'a}rio}, \citenamefont {Lee},\ and\ \citenamefont
  {Winn}}]{Apolinario:2022vzg}%
  \BibitemOpen
  \bibfield  {author} {\bibinfo {author} {\bibfnamefont {L.}~\bibnamefont
  {Apolin{\'a}rio}}, \bibinfo {author} {\bibfnamefont {Y.-J.}\ \bibnamefont
  {Lee}}, \ and\ \bibinfo {author} {\bibfnamefont {M.}~\bibnamefont {Winn}},\
  }\href {\doibase 10.1016/j.ppnp.2022.103990} {\bibfield  {journal} {\bibinfo
  {journal} {Prog. Part. Nucl. Phys.}\ }\textbf {\bibinfo {volume} {127}},\
  \bibinfo {pages} {103990} (\bibinfo {year} {2022})},\ \Eprint
  {http://arxiv.org/abs/2203.16352} {arXiv:2203.16352 [hep-ph]} \BibitemShut
  {NoStop}%
\bibitem [{\citenamefont {Sorensen}\ \emph {et~al.}(2024)\citenamefont
  {Sorensen} \emph {et~al.}}]{Sorensen:2023zkk}%
  \BibitemOpen
  \bibfield  {author} {\bibinfo {author} {\bibfnamefont {A.}~\bibnamefont
  {Sorensen}} \emph {et~al.},\ }\href {\doibase 10.1016/j.ppnp.2023.104080}
  {\bibfield  {journal} {\bibinfo  {journal} {Prog. Part. Nucl. Phys.}\
  }\textbf {\bibinfo {volume} {134}},\ \bibinfo {pages} {104080} (\bibinfo
  {year} {2024})},\ \Eprint {http://arxiv.org/abs/2301.13253} {arXiv:2301.13253
  [nucl-th]} \BibitemShut {NoStop}%
\bibitem [{\citenamefont {Arslandok}\ \emph {et~al.}(2023)\citenamefont
  {Arslandok} \emph {et~al.}}]{Arslandok:2023utm}%
  \BibitemOpen
  \bibfield  {author} {\bibinfo {author} {\bibfnamefont {M.}~\bibnamefont
  {Arslandok}} \emph {et~al.},\ }\href@noop {} {\  (\bibinfo {year} {2023})},\
  \Eprint {http://arxiv.org/abs/2303.17254} {arXiv:2303.17254 [nucl-ex]}
  \BibitemShut {NoStop}%
\bibitem [{\citenamefont {Mehtar-Tani}(2025)}]{Mehtar-Tani:2025rty}%
  \BibitemOpen
  \bibfield  {author} {\bibinfo {author} {\bibfnamefont {Y.}~\bibnamefont
  {Mehtar-Tani}},\ }\href@noop {} {\  (\bibinfo {year} {2025})},\ \Eprint
  {http://arxiv.org/abs/2509.26394} {arXiv:2509.26394 [hep-ph]} \BibitemShut
  {NoStop}%
\bibitem [{\citenamefont {Larkoski}\ \emph {et~al.}(2014)\citenamefont
  {Larkoski}, \citenamefont {Marzani}, \citenamefont {Soyez},\ and\
  \citenamefont {Thaler}}]{Larkoski:2014wba}%
  \BibitemOpen
  \bibfield  {author} {\bibinfo {author} {\bibfnamefont {A.~J.}\ \bibnamefont
  {Larkoski}}, \bibinfo {author} {\bibfnamefont {S.}~\bibnamefont {Marzani}},
  \bibinfo {author} {\bibfnamefont {G.}~\bibnamefont {Soyez}}, \ and\ \bibinfo
  {author} {\bibfnamefont {J.}~\bibnamefont {Thaler}},\ }\href {\doibase
  10.1007/JHEP05(2014)146} {\bibfield  {journal} {\bibinfo  {journal} {JHEP}\
  }\textbf {\bibinfo {volume} {05}},\ \bibinfo {pages} {146} (\bibinfo {year}
  {2014})},\ \Eprint {http://arxiv.org/abs/1402.2657} {arXiv:1402.2657
  [hep-ph]} \BibitemShut {NoStop}%
\bibitem [{\citenamefont {Krohn}\ \emph {et~al.}(2010)\citenamefont {Krohn},
  \citenamefont {Thaler},\ and\ \citenamefont {Wang}}]{Krohn:2009th}%
  \BibitemOpen
  \bibfield  {author} {\bibinfo {author} {\bibfnamefont {D.}~\bibnamefont
  {Krohn}}, \bibinfo {author} {\bibfnamefont {J.}~\bibnamefont {Thaler}}, \
  and\ \bibinfo {author} {\bibfnamefont {L.-T.}\ \bibnamefont {Wang}},\ }\href
  {\doibase 10.1007/JHEP02(2010)084} {\bibfield  {journal} {\bibinfo  {journal}
  {JHEP}\ }\textbf {\bibinfo {volume} {02}},\ \bibinfo {pages} {084} (\bibinfo
  {year} {2010})},\ \Eprint {http://arxiv.org/abs/0912.1342} {arXiv:0912.1342
  [hep-ph]} \BibitemShut {NoStop}%
\bibitem [{\citenamefont {Ellis}\ \emph {et~al.}(2009)\citenamefont {Ellis},
  \citenamefont {Vermilion},\ and\ \citenamefont {Walsh}}]{Ellis:2009su}%
  \BibitemOpen
  \bibfield  {author} {\bibinfo {author} {\bibfnamefont {S.~D.}\ \bibnamefont
  {Ellis}}, \bibinfo {author} {\bibfnamefont {C.~K.}\ \bibnamefont
  {Vermilion}}, \ and\ \bibinfo {author} {\bibfnamefont {J.~R.}\ \bibnamefont
  {Walsh}},\ }\href {\doibase 10.1103/PhysRevD.80.051501} {\bibfield  {journal}
  {\bibinfo  {journal} {Phys. Rev. D}\ }\textbf {\bibinfo {volume} {80}},\
  \bibinfo {pages} {051501} (\bibinfo {year} {2009})},\ \Eprint
  {http://arxiv.org/abs/0903.5081} {arXiv:0903.5081 [hep-ph]} \BibitemShut
  {NoStop}%
\bibitem [{\citenamefont {Butterworth}\ \emph {et~al.}(2008)\citenamefont
  {Butterworth}, \citenamefont {Davison}, \citenamefont {Rubin},\ and\
  \citenamefont {Salam}}]{Butterworth:2008iy}%
  \BibitemOpen
  \bibfield  {author} {\bibinfo {author} {\bibfnamefont {J.~M.}\ \bibnamefont
  {Butterworth}}, \bibinfo {author} {\bibfnamefont {A.~R.}\ \bibnamefont
  {Davison}}, \bibinfo {author} {\bibfnamefont {M.}~\bibnamefont {Rubin}}, \
  and\ \bibinfo {author} {\bibfnamefont {G.~P.}\ \bibnamefont {Salam}},\ }\href
  {\doibase 10.1103/PhysRevLett.100.242001} {\bibfield  {journal} {\bibinfo
  {journal} {Phys. Rev. Lett.}\ }\textbf {\bibinfo {volume} {100}},\ \bibinfo
  {pages} {242001} (\bibinfo {year} {2008})},\ \Eprint
  {http://arxiv.org/abs/0802.2470} {arXiv:0802.2470 [hep-ph]} \BibitemShut
  {NoStop}%
\bibitem [{\citenamefont {Chien}\ and\ \citenamefont
  {Vitev}(2017)}]{Chien:2016led}%
  \BibitemOpen
  \bibfield  {author} {\bibinfo {author} {\bibfnamefont {Y.-T.}\ \bibnamefont
  {Chien}}\ and\ \bibinfo {author} {\bibfnamefont {I.}~\bibnamefont {Vitev}},\
  }\href {\doibase 10.1103/PhysRevLett.119.112301} {\bibfield  {journal}
  {\bibinfo  {journal} {Phys. Rev. Lett.}\ }\textbf {\bibinfo {volume} {119}},\
  \bibinfo {pages} {112301} (\bibinfo {year} {2017})},\ \Eprint
  {http://arxiv.org/abs/1608.07283} {arXiv:1608.07283 [hep-ph]} \BibitemShut
  {NoStop}%
\bibitem [{\citenamefont {Mehtar-Tani}\ and\ \citenamefont
  {Tywoniuk}(2017)}]{Mehtar-Tani:2016aco}%
  \BibitemOpen
  \bibfield  {author} {\bibinfo {author} {\bibfnamefont {Y.}~\bibnamefont
  {Mehtar-Tani}}\ and\ \bibinfo {author} {\bibfnamefont {K.}~\bibnamefont
  {Tywoniuk}},\ }\href {\doibase 10.1007/JHEP04(2017)125} {\bibfield  {journal}
  {\bibinfo  {journal} {JHEP}\ }\textbf {\bibinfo {volume} {04}},\ \bibinfo
  {pages} {125} (\bibinfo {year} {2017})},\ \Eprint
  {http://arxiv.org/abs/1610.08930} {arXiv:1610.08930 [hep-ph]} \BibitemShut
  {NoStop}%
\bibitem [{\citenamefont {Milhano}\ \emph {et~al.}(2018)\citenamefont
  {Milhano}, \citenamefont {Wiedemann},\ and\ \citenamefont
  {Zapp}}]{Milhano:2017nzm}%
  \BibitemOpen
  \bibfield  {author} {\bibinfo {author} {\bibfnamefont {G.}~\bibnamefont
  {Milhano}}, \bibinfo {author} {\bibfnamefont {U.~A.}\ \bibnamefont
  {Wiedemann}}, \ and\ \bibinfo {author} {\bibfnamefont {K.~C.}\ \bibnamefont
  {Zapp}},\ }\href {\doibase 10.1016/j.physletb.2018.01.029} {\bibfield
  {journal} {\bibinfo  {journal} {Phys. Lett. B}\ }\textbf {\bibinfo {volume}
  {779}},\ \bibinfo {pages} {409} (\bibinfo {year} {2018})},\ \Eprint
  {http://arxiv.org/abs/1707.04142} {arXiv:1707.04142 [hep-ph]} \BibitemShut
  {NoStop}%
\bibitem [{\citenamefont {Caucal}\ \emph {et~al.}(2019)\citenamefont {Caucal},
  \citenamefont {Iancu},\ and\ \citenamefont {Soyez}}]{Caucal:2019uvr}%
  \BibitemOpen
  \bibfield  {author} {\bibinfo {author} {\bibfnamefont {P.}~\bibnamefont
  {Caucal}}, \bibinfo {author} {\bibfnamefont {E.}~\bibnamefont {Iancu}}, \
  and\ \bibinfo {author} {\bibfnamefont {G.}~\bibnamefont {Soyez}},\ }\href
  {\doibase 10.1007/JHEP10(2019)273} {\bibfield  {journal} {\bibinfo  {journal}
  {JHEP}\ }\textbf {\bibinfo {volume} {10}},\ \bibinfo {pages} {273} (\bibinfo
  {year} {2019})},\ \Eprint {http://arxiv.org/abs/1907.04866} {arXiv:1907.04866
  [hep-ph]} \BibitemShut {NoStop}%
\bibitem [{\citenamefont {Chang}\ \emph {et~al.}(2018)\citenamefont {Chang},
  \citenamefont {Cao},\ and\ \citenamefont {Qin}}]{Chang:2017gkt}%
  \BibitemOpen
  \bibfield  {author} {\bibinfo {author} {\bibfnamefont {N.-B.}\ \bibnamefont
  {Chang}}, \bibinfo {author} {\bibfnamefont {S.}~\bibnamefont {Cao}}, \ and\
  \bibinfo {author} {\bibfnamefont {G.-Y.}\ \bibnamefont {Qin}},\ }\href
  {\doibase 10.1016/j.physletb.2018.04.019} {\bibfield  {journal} {\bibinfo
  {journal} {Phys. Lett. B}\ }\textbf {\bibinfo {volume} {781}},\ \bibinfo
  {pages} {423} (\bibinfo {year} {2018})},\ \Eprint
  {http://arxiv.org/abs/1707.03767} {arXiv:1707.03767 [hep-ph]} \BibitemShut
  {NoStop}%
\bibitem [{\citenamefont {Li}\ and\ \citenamefont {Vitev}(2019)}]{Li:2017wwc}%
  \BibitemOpen
  \bibfield  {author} {\bibinfo {author} {\bibfnamefont {H.~T.}\ \bibnamefont
  {Li}}\ and\ \bibinfo {author} {\bibfnamefont {I.}~\bibnamefont {Vitev}},\
  }\href {\doibase 10.1016/j.physletb.2019.04.052} {\bibfield  {journal}
  {\bibinfo  {journal} {Phys. Lett. B}\ }\textbf {\bibinfo {volume} {793}},\
  \bibinfo {pages} {259} (\bibinfo {year} {2019})},\ \Eprint
  {http://arxiv.org/abs/1801.00008} {arXiv:1801.00008 [hep-ph]} \BibitemShut
  {NoStop}%
\bibitem [{\citenamefont {Ringer}\ \emph {et~al.}(2020)\citenamefont {Ringer},
  \citenamefont {Xiao},\ and\ \citenamefont {Yuan}}]{Ringer:2019rfk}%
  \BibitemOpen
  \bibfield  {author} {\bibinfo {author} {\bibfnamefont {F.}~\bibnamefont
  {Ringer}}, \bibinfo {author} {\bibfnamefont {B.-W.}\ \bibnamefont {Xiao}}, \
  and\ \bibinfo {author} {\bibfnamefont {F.}~\bibnamefont {Yuan}},\ }\href
  {\doibase 10.1016/j.physletb.2020.135634} {\bibfield  {journal} {\bibinfo
  {journal} {Phys. Lett. B}\ }\textbf {\bibinfo {volume} {808}},\ \bibinfo
  {pages} {135634} (\bibinfo {year} {2020})},\ \Eprint
  {http://arxiv.org/abs/1907.12541} {arXiv:1907.12541 [hep-ph]} \BibitemShut
  {NoStop}%
\bibitem [{\citenamefont {Caucal}\ \emph {et~al.}(2022)\citenamefont {Caucal},
  \citenamefont {Soto-Ontoso},\ and\ \citenamefont {Takacs}}]{Caucal:2021cfb}%
  \BibitemOpen
  \bibfield  {author} {\bibinfo {author} {\bibfnamefont {P.}~\bibnamefont
  {Caucal}}, \bibinfo {author} {\bibfnamefont {A.}~\bibnamefont {Soto-Ontoso}},
  \ and\ \bibinfo {author} {\bibfnamefont {A.}~\bibnamefont {Takacs}},\ }\href
  {\doibase 10.1103/PhysRevD.105.114046} {\bibfield  {journal} {\bibinfo
  {journal} {Phys. Rev. D}\ }\textbf {\bibinfo {volume} {105}},\ \bibinfo
  {pages} {114046} (\bibinfo {year} {2022})},\ \Eprint
  {http://arxiv.org/abs/2111.14768} {arXiv:2111.14768 [hep-ph]} \BibitemShut
  {NoStop}%
\bibitem [{\citenamefont {Cunqueiro}\ \emph {et~al.}(2023)\citenamefont
  {Cunqueiro}, \citenamefont {Napoletano},\ and\ \citenamefont
  {Soto-Ontoso}}]{Cunqueiro:2022svx}%
  \BibitemOpen
  \bibfield  {author} {\bibinfo {author} {\bibfnamefont {L.}~\bibnamefont
  {Cunqueiro}}, \bibinfo {author} {\bibfnamefont {D.}~\bibnamefont
  {Napoletano}}, \ and\ \bibinfo {author} {\bibfnamefont {A.}~\bibnamefont
  {Soto-Ontoso}},\ }\href {\doibase 10.1103/PhysRevD.107.094008} {\bibfield
  {journal} {\bibinfo  {journal} {Phys. Rev. D}\ }\textbf {\bibinfo {volume}
  {107}},\ \bibinfo {pages} {094008} (\bibinfo {year} {2023})},\ \Eprint
  {http://arxiv.org/abs/2211.11789} {arXiv:2211.11789 [hep-ph]} \BibitemShut
  {NoStop}%
\bibitem [{\citenamefont {Wang}\ \emph
  {et~al.}(2023{\natexlab{a}})\citenamefont {Wang}, \citenamefont {Kang},
  \citenamefont {Zhang}, \citenamefont {Shen}, \citenamefont {Dai},
  \citenamefont {Zhang},\ and\ \citenamefont {Wang}}]{Wang:2022yrp}%
  \BibitemOpen
  \bibfield  {author} {\bibinfo {author} {\bibfnamefont {L.}~\bibnamefont
  {Wang}}, \bibinfo {author} {\bibfnamefont {J.-W.}\ \bibnamefont {Kang}},
  \bibinfo {author} {\bibfnamefont {Q.}~\bibnamefont {Zhang}}, \bibinfo
  {author} {\bibfnamefont {S.}~\bibnamefont {Shen}}, \bibinfo {author}
  {\bibfnamefont {W.}~\bibnamefont {Dai}}, \bibinfo {author} {\bibfnamefont
  {B.-W.}\ \bibnamefont {Zhang}}, \ and\ \bibinfo {author} {\bibfnamefont
  {E.}~\bibnamefont {Wang}},\ }\href {\doibase 10.1088/0256-307X/40/3/032101}
  {\bibfield  {journal} {\bibinfo  {journal} {Chin. Phys. Lett.}\ }\textbf
  {\bibinfo {volume} {40}},\ \bibinfo {pages} {032101} (\bibinfo {year}
  {2023}{\natexlab{a}})},\ \Eprint {http://arxiv.org/abs/2211.13674}
  {arXiv:2211.13674 [nucl-th]} \BibitemShut {NoStop}%
\bibitem [{\citenamefont {Tachibana}\ \emph {et~al.}(2024)\citenamefont
  {Tachibana} \emph {et~al.}}]{JETSCAPE:2023hqn}%
  \BibitemOpen
  \bibfield  {author} {\bibinfo {author} {\bibfnamefont {Y.}~\bibnamefont
  {Tachibana}} \emph {et~al.} (\bibinfo {collaboration} {JETSCAPE}),\ }\href
  {\doibase 10.1103/PhysRevC.110.044907} {\bibfield  {journal} {\bibinfo
  {journal} {Phys. Rev. C}\ }\textbf {\bibinfo {volume} {110}},\ \bibinfo
  {pages} {044907} (\bibinfo {year} {2024})},\ \Eprint
  {http://arxiv.org/abs/2301.02485} {arXiv:2301.02485 [hep-ph]} \BibitemShut
  {NoStop}%
\bibitem [{\citenamefont {Tachibana}\ \emph {et~al.}(2026)\citenamefont
  {Tachibana} \emph {et~al.}}]{Tachibana:2025rcx}%
  \BibitemOpen
  \bibfield  {author} {\bibinfo {author} {\bibfnamefont {Y.}~\bibnamefont
  {Tachibana}} \emph {et~al.},\ }\href {\doibase 10.1103/7qxy-fz41} {\bibfield
  {journal} {\bibinfo  {journal} {Phys. Rev. C}\ }\textbf {\bibinfo {volume}
  {113}},\ \bibinfo {pages} {034910} (\bibinfo {year} {2026})},\ \Eprint
  {http://arxiv.org/abs/2503.23693} {arXiv:2503.23693 [hep-ph]} \BibitemShut
  {NoStop}%
\bibitem [{\citenamefont {Acharya}\ \emph {et~al.}(2025)\citenamefont {Acharya}
  \emph {et~al.}}]{ALICE:2024fip}%
  \BibitemOpen
  \bibfield  {author} {\bibinfo {author} {\bibfnamefont {S.}~\bibnamefont
  {Acharya}} \emph {et~al.} (\bibinfo {collaboration} {ALICE}),\ }\href
  {\doibase 10.1103/PhysRevLett.135.031901} {\bibfield  {journal} {\bibinfo
  {journal} {Phys. Rev. Lett.}\ }\textbf {\bibinfo {volume} {135}},\ \bibinfo
  {pages} {031901} (\bibinfo {year} {2025})},\ \Eprint
  {http://arxiv.org/abs/2409.12837} {arXiv:2409.12837 [nucl-ex]} \BibitemShut
  {NoStop}%
\bibitem [{\citenamefont {Baier}\ \emph {et~al.}(2001)\citenamefont {Baier},
  \citenamefont {Dokshitzer}, \citenamefont {Mueller},\ and\ \citenamefont
  {Schiff}}]{Baier:2001yt}%
  \BibitemOpen
  \bibfield  {author} {\bibinfo {author} {\bibfnamefont {R.}~\bibnamefont
  {Baier}}, \bibinfo {author} {\bibfnamefont {Y.~L.}\ \bibnamefont
  {Dokshitzer}}, \bibinfo {author} {\bibfnamefont {A.~H.}\ \bibnamefont
  {Mueller}}, \ and\ \bibinfo {author} {\bibfnamefont {D.}~\bibnamefont
  {Schiff}},\ }\href {\doibase 10.1088/1126-6708/2001/09/033} {\bibfield
  {journal} {\bibinfo  {journal} {JHEP}\ }\textbf {\bibinfo {volume} {09}},\
  \bibinfo {pages} {033} (\bibinfo {year} {2001})},\ \Eprint
  {http://arxiv.org/abs/hep-ph/0106347} {arXiv:hep-ph/0106347} \BibitemShut
  {NoStop}%
\bibitem [{\citenamefont {Renk}(2013)}]{Renk:2012ve}%
  \BibitemOpen
  \bibfield  {author} {\bibinfo {author} {\bibfnamefont {T.}~\bibnamefont
  {Renk}},\ }\href {\doibase 10.1103/PhysRevC.88.054902} {\bibfield  {journal}
  {\bibinfo  {journal} {Phys. Rev. C}\ }\textbf {\bibinfo {volume} {88}},\
  \bibinfo {pages} {054902} (\bibinfo {year} {2013})},\ \Eprint
  {http://arxiv.org/abs/1212.0646} {arXiv:1212.0646 [hep-ph]} \BibitemShut
  {NoStop}%
\bibitem [{\citenamefont {Wang}\ \emph
  {et~al.}(2021{\natexlab{a}})\citenamefont {Wang}, \citenamefont {Kang},
  \citenamefont {Dai}, \citenamefont {Zhang},\ and\ \citenamefont
  {Wang}}]{Wang:2021jgm}%
  \BibitemOpen
  \bibfield  {author} {\bibinfo {author} {\bibfnamefont {S.}~\bibnamefont
  {Wang}}, \bibinfo {author} {\bibfnamefont {J.-W.}\ \bibnamefont {Kang}},
  \bibinfo {author} {\bibfnamefont {W.}~\bibnamefont {Dai}}, \bibinfo {author}
  {\bibfnamefont {B.-W.}\ \bibnamefont {Zhang}}, \ and\ \bibinfo {author}
  {\bibfnamefont {E.}~\bibnamefont {Wang}},\ }\href {\doibase
  10.1140/epja/s10050-022-00785-9} {\bibfield  {journal} {\bibinfo  {journal}
  {Eur. Phys. J. A}\ }\textbf {\bibinfo {volume} {58}},\ \bibinfo {pages} {149}
  (\bibinfo {year} {2021}{\natexlab{a}})},\ \Eprint
  {http://arxiv.org/abs/2107.12000} {arXiv:2107.12000 [nucl-th]} \BibitemShut
  {NoStop}%
\bibitem [{\citenamefont {Brewer}\ \emph {et~al.}(2022)\citenamefont {Brewer},
  \citenamefont {Brodsky},\ and\ \citenamefont {Rajagopal}}]{Brewer:2021hmh}%
  \BibitemOpen
  \bibfield  {author} {\bibinfo {author} {\bibfnamefont {J.}~\bibnamefont
  {Brewer}}, \bibinfo {author} {\bibfnamefont {Q.}~\bibnamefont {Brodsky}}, \
  and\ \bibinfo {author} {\bibfnamefont {K.}~\bibnamefont {Rajagopal}},\ }\href
  {\doibase 10.1007/JHEP02(2022)175} {\bibfield  {journal} {\bibinfo  {journal}
  {JHEP}\ }\textbf {\bibinfo {volume} {02}},\ \bibinfo {pages} {175} (\bibinfo
  {year} {2022})},\ \Eprint {http://arxiv.org/abs/2110.13159} {arXiv:2110.13159
  [hep-ph]} \BibitemShut {NoStop}%
\bibitem [{\citenamefont {Acharya}\ \emph {et~al.}(2023)\citenamefont {Acharya}
  \emph {et~al.}}]{ALICE:2023dwg}%
  \BibitemOpen
  \bibfield  {author} {\bibinfo {author} {\bibfnamefont {S.}~\bibnamefont
  {Acharya}} \emph {et~al.} (\bibinfo {collaboration} {ALICE}),\ }\href@noop {}
  {\  (\bibinfo {year} {2023})},\ \Eprint {http://arxiv.org/abs/2303.13347}
  {arXiv:2303.13347 [nucl-ex]} \BibitemShut {NoStop}%
\bibitem [{\citenamefont {Acharya}\ \emph {et~al.}(2018)\citenamefont {Acharya}
  \emph {et~al.}}]{ALICE:2018dxf}%
  \BibitemOpen
  \bibfield  {author} {\bibinfo {author} {\bibfnamefont {S.}~\bibnamefont
  {Acharya}} \emph {et~al.} (\bibinfo {collaboration} {ALICE}),\ }\href
  {\doibase 10.1007/JHEP10(2018)139} {\bibfield  {journal} {\bibinfo  {journal}
  {JHEP}\ }\textbf {\bibinfo {volume} {10}},\ \bibinfo {pages} {139} (\bibinfo
  {year} {2018})},\ \Eprint {http://arxiv.org/abs/1807.06854} {arXiv:1807.06854
  [nucl-ex]} \BibitemShut {NoStop}%
\bibitem [{\citenamefont {Yan}\ \emph {et~al.}(2021)\citenamefont {Yan},
  \citenamefont {Chen}, \citenamefont {Dai}, \citenamefont {Zhang},\ and\
  \citenamefont {Wang}}]{Yan:2020zrz}%
  \BibitemOpen
  \bibfield  {author} {\bibinfo {author} {\bibfnamefont {J.}~\bibnamefont
  {Yan}}, \bibinfo {author} {\bibfnamefont {S.-Y.}\ \bibnamefont {Chen}},
  \bibinfo {author} {\bibfnamefont {W.}~\bibnamefont {Dai}}, \bibinfo {author}
  {\bibfnamefont {B.-W.}\ \bibnamefont {Zhang}}, \ and\ \bibinfo {author}
  {\bibfnamefont {E.}~\bibnamefont {Wang}},\ }\href {\doibase
  10.1088/1674-1137/abca2b} {\bibfield  {journal} {\bibinfo  {journal} {Chin.
  Phys. C}\ }\textbf {\bibinfo {volume} {45}},\ \bibinfo {pages} {024102}
  (\bibinfo {year} {2021})},\ \Eprint {http://arxiv.org/abs/2005.01093}
  {arXiv:2005.01093 [hep-ph]} \BibitemShut {NoStop}%
\bibitem [{\citenamefont {Hayrapetyan}\ \emph {et~al.}(2025)\citenamefont
  {Hayrapetyan} \emph {et~al.}}]{CMS:2024zjn}%
  \BibitemOpen
  \bibfield  {author} {\bibinfo {author} {\bibfnamefont {A.}~\bibnamefont
  {Hayrapetyan}} \emph {et~al.} (\bibinfo {collaboration} {CMS}),\ }\href
  {\doibase 10.1016/j.physletb.2024.139088} {\bibfield  {journal} {\bibinfo
  {journal} {Phys. Lett. B}\ }\textbf {\bibinfo {volume} {861}},\ \bibinfo
  {pages} {139088} (\bibinfo {year} {2025})},\ \Eprint
  {http://arxiv.org/abs/2405.02737} {arXiv:2405.02737 [nucl-ex]} \BibitemShut
  {NoStop}%
\bibitem [{\citenamefont {Wang}\ \emph {et~al.}(2024)\citenamefont {Wang},
  \citenamefont {Li}, \citenamefont {Kang},\ and\ \citenamefont
  {Zhang}}]{Wang:2024plm}%
  \BibitemOpen
  \bibfield  {author} {\bibinfo {author} {\bibfnamefont {S.}~\bibnamefont
  {Wang}}, \bibinfo {author} {\bibfnamefont {Y.}~\bibnamefont {Li}}, \bibinfo
  {author} {\bibfnamefont {J.-W.}\ \bibnamefont {Kang}}, \ and\ \bibinfo
  {author} {\bibfnamefont {B.-W.}\ \bibnamefont {Zhang}},\ }\href@noop {} {\
  (\bibinfo {year} {2024})},\ \Eprint {http://arxiv.org/abs/2408.10924}
  {arXiv:2408.10924 [hep-ph]} \BibitemShut {NoStop}%
\bibitem [{\citenamefont {Aboona}\ \emph {et~al.}(2025)\citenamefont {Aboona}
  \emph {et~al.}}]{STAR:2023pal}%
  \BibitemOpen
  \bibfield  {author} {\bibinfo {author} {\bibfnamefont {B.~E.}\ \bibnamefont
  {Aboona}} \emph {et~al.} (\bibinfo {collaboration} {STAR}),\ }\href {\doibase
  10.1103/PhysRevLett.134.232301} {\bibfield  {journal} {\bibinfo  {journal}
  {Phys. Rev. Lett.}\ }\textbf {\bibinfo {volume} {134}},\ \bibinfo {pages}
  {232301} (\bibinfo {year} {2025})},\ \Eprint
  {http://arxiv.org/abs/2309.00156} {arXiv:2309.00156 [nucl-ex]} \BibitemShut
  {NoStop}%
\bibitem [{\citenamefont {Acharya}\ \emph {et~al.}(2024)\citenamefont {Acharya}
  \emph {et~al.}}]{ALICE:2023qve}%
  \BibitemOpen
  \bibfield  {author} {\bibinfo {author} {\bibfnamefont {S.}~\bibnamefont
  {Acharya}} \emph {et~al.} (\bibinfo {collaboration} {ALICE}),\ }\href
  {\doibase 10.1103/PhysRevLett.133.022301} {\bibfield  {journal} {\bibinfo
  {journal} {Phys. Rev. Lett.}\ }\textbf {\bibinfo {volume} {133}},\ \bibinfo
  {pages} {022301} (\bibinfo {year} {2024})},\ \Eprint
  {http://arxiv.org/abs/2308.16131} {arXiv:2308.16131 [nucl-ex]} \BibitemShut
  {NoStop}%
\bibitem [{\citenamefont {Mehtar-Tani}\ \emph {et~al.}(2020)\citenamefont
  {Mehtar-Tani}, \citenamefont {Soto-Ontoso},\ and\ \citenamefont
  {Tywoniuk}}]{Mehtar-Tani:2019rrk}%
  \BibitemOpen
  \bibfield  {author} {\bibinfo {author} {\bibfnamefont {Y.}~\bibnamefont
  {Mehtar-Tani}}, \bibinfo {author} {\bibfnamefont {A.}~\bibnamefont
  {Soto-Ontoso}}, \ and\ \bibinfo {author} {\bibfnamefont {K.}~\bibnamefont
  {Tywoniuk}},\ }\href {\doibase 10.1103/PhysRevD.101.034004} {\bibfield
  {journal} {\bibinfo  {journal} {Phys. Rev. D}\ }\textbf {\bibinfo {volume}
  {101}},\ \bibinfo {pages} {034004} (\bibinfo {year} {2020})},\ \Eprint
  {http://arxiv.org/abs/1911.00375} {arXiv:1911.00375 [hep-ph]} \BibitemShut
  {NoStop}%
\bibitem [{\citenamefont {Cacciari}\ \emph {et~al.}(2008)\citenamefont
  {Cacciari}, \citenamefont {Salam},\ and\ \citenamefont
  {Soyez}}]{Cacciari:2008gp}%
  \BibitemOpen
  \bibfield  {author} {\bibinfo {author} {\bibfnamefont {M.}~\bibnamefont
  {Cacciari}}, \bibinfo {author} {\bibfnamefont {G.~P.}\ \bibnamefont {Salam}},
  \ and\ \bibinfo {author} {\bibfnamefont {G.}~\bibnamefont {Soyez}},\ }\href
  {\doibase 10.1088/1126-6708/2008/04/063} {\bibfield  {journal} {\bibinfo
  {journal} {JHEP}\ }\textbf {\bibinfo {volume} {04}},\ \bibinfo {pages} {063}
  (\bibinfo {year} {2008})},\ \Eprint {http://arxiv.org/abs/0802.1189}
  {arXiv:0802.1189 [hep-ph]} \BibitemShut {NoStop}%
\bibitem [{\citenamefont {Dokshitzer}\ \emph {et~al.}(1997)\citenamefont
  {Dokshitzer}, \citenamefont {Leder}, \citenamefont {Moretti},\ and\
  \citenamefont {Webber}}]{Dokshitzer:1997in}%
  \BibitemOpen
  \bibfield  {author} {\bibinfo {author} {\bibfnamefont {Y.~L.}\ \bibnamefont
  {Dokshitzer}}, \bibinfo {author} {\bibfnamefont {G.~D.}\ \bibnamefont
  {Leder}}, \bibinfo {author} {\bibfnamefont {S.}~\bibnamefont {Moretti}}, \
  and\ \bibinfo {author} {\bibfnamefont {B.~R.}\ \bibnamefont {Webber}},\
  }\href {\doibase 10.1088/1126-6708/1997/08/001} {\bibfield  {journal}
  {\bibinfo  {journal} {JHEP}\ }\textbf {\bibinfo {volume} {08}},\ \bibinfo
  {pages} {001} (\bibinfo {year} {1997})},\ \Eprint
  {http://arxiv.org/abs/hep-ph/9707323} {arXiv:hep-ph/9707323} \BibitemShut
  {NoStop}%
\bibitem [{\citenamefont {Dasgupta}\ \emph {et~al.}(2013)\citenamefont
  {Dasgupta}, \citenamefont {Fregoso}, \citenamefont {Marzani},\ and\
  \citenamefont {Salam}}]{Dasgupta:2013ihk}%
  \BibitemOpen
  \bibfield  {author} {\bibinfo {author} {\bibfnamefont {M.}~\bibnamefont
  {Dasgupta}}, \bibinfo {author} {\bibfnamefont {A.}~\bibnamefont {Fregoso}},
  \bibinfo {author} {\bibfnamefont {S.}~\bibnamefont {Marzani}}, \ and\
  \bibinfo {author} {\bibfnamefont {G.~P.}\ \bibnamefont {Salam}},\ }\href
  {\doibase 10.1007/JHEP09(2013)029} {\bibfield  {journal} {\bibinfo  {journal}
  {JHEP}\ }\textbf {\bibinfo {volume} {09}},\ \bibinfo {pages} {029} (\bibinfo
  {year} {2013})},\ \Eprint {http://arxiv.org/abs/1307.0007} {arXiv:1307.0007
  [hep-ph]} \BibitemShut {NoStop}%
\bibitem [{\citenamefont {Sj{\"o}strand}\ \emph {et~al.}(2015)\citenamefont
  {Sj{\"o}strand}, \citenamefont {Ask}, \citenamefont {Christiansen},
  \citenamefont {Corke}, \citenamefont {Desai}, \citenamefont {Ilten},
  \citenamefont {Mrenna}, \citenamefont {Prestel}, \citenamefont {Rasmussen},\
  and\ \citenamefont {Skands}}]{Sjostrand:2014zea}%
  \BibitemOpen
  \bibfield  {author} {\bibinfo {author} {\bibfnamefont {T.}~\bibnamefont
  {Sj{\"o}strand}}, \bibinfo {author} {\bibfnamefont {S.}~\bibnamefont {Ask}},
  \bibinfo {author} {\bibfnamefont {J.~R.}\ \bibnamefont {Christiansen}},
  \bibinfo {author} {\bibfnamefont {R.}~\bibnamefont {Corke}}, \bibinfo
  {author} {\bibfnamefont {N.}~\bibnamefont {Desai}}, \bibinfo {author}
  {\bibfnamefont {P.}~\bibnamefont {Ilten}}, \bibinfo {author} {\bibfnamefont
  {S.}~\bibnamefont {Mrenna}}, \bibinfo {author} {\bibfnamefont
  {S.}~\bibnamefont {Prestel}}, \bibinfo {author} {\bibfnamefont {C.~O.}\
  \bibnamefont {Rasmussen}}, \ and\ \bibinfo {author} {\bibfnamefont {P.~Z.}\
  \bibnamefont {Skands}},\ }\href {\doibase 10.1016/j.cpc.2015.01.024}
  {\bibfield  {journal} {\bibinfo  {journal} {Comput. Phys. Commun.}\ }\textbf
  {\bibinfo {volume} {191}},\ \bibinfo {pages} {159} (\bibinfo {year}
  {2015})},\ \Eprint {http://arxiv.org/abs/1410.3012} {arXiv:1410.3012
  [hep-ph]} \BibitemShut {NoStop}%
\bibitem [{\citenamefont {Skands}\ \emph {et~al.}(2014)\citenamefont {Skands},
  \citenamefont {Carrazza},\ and\ \citenamefont {Rojo}}]{Skands:2014pea}%
  \BibitemOpen
  \bibfield  {author} {\bibinfo {author} {\bibfnamefont {P.}~\bibnamefont
  {Skands}}, \bibinfo {author} {\bibfnamefont {S.}~\bibnamefont {Carrazza}}, \
  and\ \bibinfo {author} {\bibfnamefont {J.}~\bibnamefont {Rojo}},\ }\href
  {\doibase 10.1140/epjc/s10052-014-3024-y} {\bibfield  {journal} {\bibinfo
  {journal} {Eur. Phys. J. C}\ }\textbf {\bibinfo {volume} {74}},\ \bibinfo
  {pages} {3024} (\bibinfo {year} {2014})},\ \Eprint
  {http://arxiv.org/abs/1404.5630} {arXiv:1404.5630 [hep-ph]} \BibitemShut
  {NoStop}%
\bibitem [{\citenamefont {Wang}\ \emph
  {et~al.}(2025{\natexlab{a}})\citenamefont {Wang}, \citenamefont {Li},
  \citenamefont {Shen}, \citenamefont {Zhang},\ and\ \citenamefont
  {Wang}}]{Wang:2023udp}%
  \BibitemOpen
  \bibfield  {author} {\bibinfo {author} {\bibfnamefont {S.}~\bibnamefont
  {Wang}}, \bibinfo {author} {\bibfnamefont {Y.}~\bibnamefont {Li}}, \bibinfo
  {author} {\bibfnamefont {S.}~\bibnamefont {Shen}}, \bibinfo {author}
  {\bibfnamefont {B.-W.}\ \bibnamefont {Zhang}}, \ and\ \bibinfo {author}
  {\bibfnamefont {E.}~\bibnamefont {Wang}},\ }\href {\doibase
  10.1103/PhysRevC.111.034912} {\bibfield  {journal} {\bibinfo  {journal}
  {Phys. Rev. C}\ }\textbf {\bibinfo {volume} {111}},\ \bibinfo {pages}
  {034912} (\bibinfo {year} {2025}{\natexlab{a}})},\ \Eprint
  {http://arxiv.org/abs/2308.14538} {arXiv:2308.14538 [hep-ph]} \BibitemShut
  {NoStop}%
\bibitem [{\citenamefont {Li}\ \emph {et~al.}(2024)\citenamefont {Li},
  \citenamefont {Shen}, \citenamefont {Wang},\ and\ \citenamefont
  {Zhang}}]{Li:2024uzk}%
  \BibitemOpen
  \bibfield  {author} {\bibinfo {author} {\bibfnamefont {Y.}~\bibnamefont
  {Li}}, \bibinfo {author} {\bibfnamefont {S.}~\bibnamefont {Shen}}, \bibinfo
  {author} {\bibfnamefont {S.}~\bibnamefont {Wang}}, \ and\ \bibinfo {author}
  {\bibfnamefont {B.-W.}\ \bibnamefont {Zhang}},\ }\href {\doibase
  10.1007/s41365-024-01482-6} {\bibfield  {journal} {\bibinfo  {journal} {Nucl.
  Sci. Tech.}\ }\textbf {\bibinfo {volume} {35}},\ \bibinfo {pages} {113}
  (\bibinfo {year} {2024})},\ \Eprint {http://arxiv.org/abs/2401.01706}
  {arXiv:2401.01706 [hep-ph]} \BibitemShut {NoStop}%
\bibitem [{\citenamefont {Wang}\ \emph
  {et~al.}(2023{\natexlab{b}})\citenamefont {Wang}, \citenamefont {Dai},
  \citenamefont {Zhang},\ and\ \citenamefont {Wang}}]{Wang:2020qwe}%
  \BibitemOpen
  \bibfield  {author} {\bibinfo {author} {\bibfnamefont {S.}~\bibnamefont
  {Wang}}, \bibinfo {author} {\bibfnamefont {W.}~\bibnamefont {Dai}}, \bibinfo
  {author} {\bibfnamefont {B.-W.}\ \bibnamefont {Zhang}}, \ and\ \bibinfo
  {author} {\bibfnamefont {E.}~\bibnamefont {Wang}},\ }\href {\doibase
  10.1088/1674-1137/acc1ca} {\bibfield  {journal} {\bibinfo  {journal} {Chin.
  Phys. C}\ }\textbf {\bibinfo {volume} {47}},\ \bibinfo {pages} {054102}
  (\bibinfo {year} {2023}{\natexlab{b}})},\ \Eprint
  {http://arxiv.org/abs/2005.07018} {arXiv:2005.07018 [hep-ph]} \BibitemShut
  {NoStop}%
\bibitem [{\citenamefont {Wang}\ \emph
  {et~al.}(2025{\natexlab{b}})\citenamefont {Wang}, \citenamefont {Li},
  \citenamefont {Li}, \citenamefont {Zhang},\ and\ \citenamefont
  {Wang}}]{Wang:2024yag}%
  \BibitemOpen
  \bibfield  {author} {\bibinfo {author} {\bibfnamefont {S.}~\bibnamefont
  {Wang}}, \bibinfo {author} {\bibfnamefont {S.}~\bibnamefont {Li}}, \bibinfo
  {author} {\bibfnamefont {Y.}~\bibnamefont {Li}}, \bibinfo {author}
  {\bibfnamefont {B.-W.}\ \bibnamefont {Zhang}}, \ and\ \bibinfo {author}
  {\bibfnamefont {E.}~\bibnamefont {Wang}},\ }\href {\doibase
  10.1088/1674-1137/adb385} {\bibfield  {journal} {\bibinfo  {journal} {Chin.
  Phys. C}\ }\textbf {\bibinfo {volume} {49}},\ \bibinfo {pages} {064101}
  (\bibinfo {year} {2025}{\natexlab{b}})},\ \Eprint
  {http://arxiv.org/abs/2410.21834} {arXiv:2410.21834 [hep-ph]} \BibitemShut
  {NoStop}%
\bibitem [{\citenamefont {Wang}\ \emph {et~al.}(2019)\citenamefont {Wang},
  \citenamefont {Dai}, \citenamefont {Zhang},\ and\ \citenamefont
  {Wang}}]{Wang:2019xey}%
  \BibitemOpen
  \bibfield  {author} {\bibinfo {author} {\bibfnamefont {S.}~\bibnamefont
  {Wang}}, \bibinfo {author} {\bibfnamefont {W.}~\bibnamefont {Dai}}, \bibinfo
  {author} {\bibfnamefont {B.-W.}\ \bibnamefont {Zhang}}, \ and\ \bibinfo
  {author} {\bibfnamefont {E.}~\bibnamefont {Wang}},\ }\href {\doibase
  10.1140/epjc/s10052-019-7312-4} {\bibfield  {journal} {\bibinfo  {journal}
  {Eur. Phys. J. C}\ }\textbf {\bibinfo {volume} {79}},\ \bibinfo {pages} {789}
  (\bibinfo {year} {2019})},\ \Eprint {http://arxiv.org/abs/1906.01499}
  {arXiv:1906.01499 [nucl-th]} \BibitemShut {NoStop}%
\bibitem [{\citenamefont {Wang}\ \emph
  {et~al.}(2023{\natexlab{c}})\citenamefont {Wang}, \citenamefont {Dai},
  \citenamefont {Wang}, \citenamefont {Wang},\ and\ \citenamefont
  {Zhang}}]{Wang:2023eer}%
  \BibitemOpen
  \bibfield  {author} {\bibinfo {author} {\bibfnamefont {S.}~\bibnamefont
  {Wang}}, \bibinfo {author} {\bibfnamefont {W.}~\bibnamefont {Dai}}, \bibinfo
  {author} {\bibfnamefont {E.}~\bibnamefont {Wang}}, \bibinfo {author}
  {\bibfnamefont {X.-N.}\ \bibnamefont {Wang}}, \ and\ \bibinfo {author}
  {\bibfnamefont {B.-W.}\ \bibnamefont {Zhang}},\ }\href {\doibase
  10.3390/sym15030727} {\bibfield  {journal} {\bibinfo  {journal} {Symmetry}\
  }\textbf {\bibinfo {volume} {15}},\ \bibinfo {pages} {727} (\bibinfo {year}
  {2023}{\natexlab{c}})},\ \Eprint {http://arxiv.org/abs/2303.14660}
  {arXiv:2303.14660 [nucl-th]} \BibitemShut {NoStop}%
\bibitem [{\citenamefont {Wang}\ \emph
  {et~al.}(2021{\natexlab{b}})\citenamefont {Wang}, \citenamefont {Dai},
  \citenamefont {Zhang},\ and\ \citenamefont {Wang}}]{Wang:2020ukj}%
  \BibitemOpen
  \bibfield  {author} {\bibinfo {author} {\bibfnamefont {S.}~\bibnamefont
  {Wang}}, \bibinfo {author} {\bibfnamefont {W.}~\bibnamefont {Dai}}, \bibinfo
  {author} {\bibfnamefont {B.-W.}\ \bibnamefont {Zhang}}, \ and\ \bibinfo
  {author} {\bibfnamefont {E.}~\bibnamefont {Wang}},\ }\href {\doibase
  10.1088/1674-1137/abf4f5} {\bibfield  {journal} {\bibinfo  {journal} {Chin.
  Phys. C}\ }\textbf {\bibinfo {volume} {45}},\ \bibinfo {pages} {064105}
  (\bibinfo {year} {2021}{\natexlab{b}})},\ \Eprint
  {http://arxiv.org/abs/2012.13935} {arXiv:2012.13935 [nucl-th]} \BibitemShut
  {NoStop}%
\bibitem [{\citenamefont {Li}\ \emph {et~al.}(2023)\citenamefont {Li},
  \citenamefont {Wang},\ and\ \citenamefont {Zhang}}]{Li:2022tcr}%
  \BibitemOpen
  \bibfield  {author} {\bibinfo {author} {\bibfnamefont {Y.}~\bibnamefont
  {Li}}, \bibinfo {author} {\bibfnamefont {S.}~\bibnamefont {Wang}}, \ and\
  \bibinfo {author} {\bibfnamefont {B.-W.}\ \bibnamefont {Zhang}},\ }\href
  {\doibase 10.1103/PhysRevC.108.024905} {\bibfield  {journal} {\bibinfo
  {journal} {Phys. Rev. C}\ }\textbf {\bibinfo {volume} {108}},\ \bibinfo
  {pages} {024905} (\bibinfo {year} {2023})},\ \Eprint
  {http://arxiv.org/abs/2209.00548} {arXiv:2209.00548 [hep-ph]} \BibitemShut
  {NoStop}%
\bibitem [{\citenamefont {Li}\ \emph {et~al.}(2025)\citenamefont {Li},
  \citenamefont {Chen}, \citenamefont {Kong}, \citenamefont {Wang},\ and\
  \citenamefont {Zhang}}]{Li:2024pfi}%
  \BibitemOpen
  \bibfield  {author} {\bibinfo {author} {\bibfnamefont {Y.}~\bibnamefont
  {Li}}, \bibinfo {author} {\bibfnamefont {S.-Y.}\ \bibnamefont {Chen}},
  \bibinfo {author} {\bibfnamefont {W.-X.}\ \bibnamefont {Kong}}, \bibinfo
  {author} {\bibfnamefont {S.}~\bibnamefont {Wang}}, \ and\ \bibinfo {author}
  {\bibfnamefont {B.-W.}\ \bibnamefont {Zhang}},\ }\href {\doibase
  10.1088/0256-307X/42/1/011201} {\bibfield  {journal} {\bibinfo  {journal}
  {Chin. Phys. Lett.}\ }\textbf {\bibinfo {volume} {42}},\ \bibinfo {pages}
  {011201} (\bibinfo {year} {2025})},\ \Eprint
  {http://arxiv.org/abs/2409.12742} {arXiv:2409.12742 [hep-ph]} \BibitemShut
  {NoStop}%
\bibitem [{\citenamefont {Guo}\ and\ \citenamefont {Wang}(2000)}]{Guo:2000nz}%
  \BibitemOpen
  \bibfield  {author} {\bibinfo {author} {\bibfnamefont {X.-f.}\ \bibnamefont
  {Guo}}\ and\ \bibinfo {author} {\bibfnamefont {X.-N.}\ \bibnamefont {Wang}},\
  }\href {\doibase 10.1103/PhysRevLett.85.3591} {\bibfield  {journal} {\bibinfo
   {journal} {Phys. Rev. Lett.}\ }\textbf {\bibinfo {volume} {85}},\ \bibinfo
  {pages} {3591} (\bibinfo {year} {2000})},\ \Eprint
  {http://arxiv.org/abs/hep-ph/0005044} {arXiv:hep-ph/0005044} \BibitemShut
  {NoStop}%
\bibitem [{\citenamefont {Zhang}\ and\ \citenamefont
  {Wang}(2003)}]{Zhang:2003yn}%
  \BibitemOpen
  \bibfield  {author} {\bibinfo {author} {\bibfnamefont {B.-W.}\ \bibnamefont
  {Zhang}}\ and\ \bibinfo {author} {\bibfnamefont {X.-N.}\ \bibnamefont
  {Wang}},\ }\href {\doibase 10.1016/S0375-9474(03)01003-0} {\bibfield
  {journal} {\bibinfo  {journal} {Nucl. Phys. A}\ }\textbf {\bibinfo {volume}
  {720}},\ \bibinfo {pages} {429} (\bibinfo {year} {2003})},\ \Eprint
  {http://arxiv.org/abs/hep-ph/0301195} {arXiv:hep-ph/0301195} \BibitemShut
  {NoStop}%
\bibitem [{\citenamefont {Zhang}\ \emph {et~al.}(2004)\citenamefont {Zhang},
  \citenamefont {Wang},\ and\ \citenamefont {Wang}}]{Zhang:2003wk}%
  \BibitemOpen
  \bibfield  {author} {\bibinfo {author} {\bibfnamefont {B.-W.}\ \bibnamefont
  {Zhang}}, \bibinfo {author} {\bibfnamefont {E.}~\bibnamefont {Wang}}, \ and\
  \bibinfo {author} {\bibfnamefont {X.-N.}\ \bibnamefont {Wang}},\ }\href
  {\doibase 10.1103/PhysRevLett.93.072301} {\bibfield  {journal} {\bibinfo
  {journal} {Phys. Rev. Lett.}\ }\textbf {\bibinfo {volume} {93}},\ \bibinfo
  {pages} {072301} (\bibinfo {year} {2004})},\ \Eprint
  {http://arxiv.org/abs/nucl-th/0309040} {arXiv:nucl-th/0309040} \BibitemShut
  {NoStop}%
\bibitem [{\citenamefont {Majumder}(2012)}]{Majumder:2009ge}%
  \BibitemOpen
  \bibfield  {author} {\bibinfo {author} {\bibfnamefont {A.}~\bibnamefont
  {Majumder}},\ }\href {\doibase 10.1103/PhysRevD.85.014023} {\bibfield
  {journal} {\bibinfo  {journal} {Phys. Rev. D}\ }\textbf {\bibinfo {volume}
  {85}},\ \bibinfo {pages} {014023} (\bibinfo {year} {2012})},\ \Eprint
  {http://arxiv.org/abs/0912.2987} {arXiv:0912.2987 [nucl-th]} \BibitemShut
  {NoStop}%
\bibitem [{\citenamefont {Deng}\ and\ \citenamefont
  {Wang}(2010)}]{Deng:2009ncl}%
  \BibitemOpen
  \bibfield  {author} {\bibinfo {author} {\bibfnamefont {W.-t.}\ \bibnamefont
  {Deng}}\ and\ \bibinfo {author} {\bibfnamefont {X.-N.}\ \bibnamefont
  {Wang}},\ }\href {\doibase 10.1103/PhysRevC.81.024902} {\bibfield  {journal}
  {\bibinfo  {journal} {Phys. Rev. C}\ }\textbf {\bibinfo {volume} {81}},\
  \bibinfo {pages} {024902} (\bibinfo {year} {2010})},\ \Eprint
  {http://arxiv.org/abs/0910.3403} {arXiv:0910.3403 [hep-ph]} \BibitemShut
  {NoStop}%
\bibitem [{\citenamefont {He}\ \emph {et~al.}(2015)\citenamefont {He},
  \citenamefont {Luo}, \citenamefont {Wang},\ and\ \citenamefont
  {Zhu}}]{He:2015pra}%
  \BibitemOpen
  \bibfield  {author} {\bibinfo {author} {\bibfnamefont {Y.}~\bibnamefont
  {He}}, \bibinfo {author} {\bibfnamefont {T.}~\bibnamefont {Luo}}, \bibinfo
  {author} {\bibfnamefont {X.-N.}\ \bibnamefont {Wang}}, \ and\ \bibinfo
  {author} {\bibfnamefont {Y.}~\bibnamefont {Zhu}},\ }\href {\doibase
  10.1103/PhysRevC.91.054908} {\bibfield  {journal} {\bibinfo  {journal} {Phys.
  Rev. C}\ }\textbf {\bibinfo {volume} {91}},\ \bibinfo {pages} {054908}
  (\bibinfo {year} {2015})},\ \bibinfo {note} {[Erratum: Phys.Rev.C 97, 019902
  (2018)]},\ \Eprint {http://arxiv.org/abs/1503.03313} {arXiv:1503.03313
  [nucl-th]} \BibitemShut {NoStop}%
\bibitem [{\citenamefont {Wang}\ \emph {et~al.}(1995)\citenamefont {Wang},
  \citenamefont {Gyulassy},\ and\ \citenamefont {Plumer}}]{Wang:1994fx}%
  \BibitemOpen
  \bibfield  {author} {\bibinfo {author} {\bibfnamefont {X.-N.}\ \bibnamefont
  {Wang}}, \bibinfo {author} {\bibfnamefont {M.}~\bibnamefont {Gyulassy}}, \
  and\ \bibinfo {author} {\bibfnamefont {M.}~\bibnamefont {Plumer}},\ }\href
  {\doibase 10.1103/PhysRevD.51.3436} {\bibfield  {journal} {\bibinfo
  {journal} {Phys. Rev. D}\ }\textbf {\bibinfo {volume} {51}},\ \bibinfo
  {pages} {3436} (\bibinfo {year} {1995})},\ \Eprint
  {http://arxiv.org/abs/hep-ph/9408344} {arXiv:hep-ph/9408344} \BibitemShut
  {NoStop}%
\bibitem [{\citenamefont {Chen}\ \emph {et~al.}(2010)\citenamefont {Chen},
  \citenamefont {Greiner}, \citenamefont {Wang}, \citenamefont {Wang},\ and\
  \citenamefont {Xu}}]{Chen:2010te}%
  \BibitemOpen
  \bibfield  {author} {\bibinfo {author} {\bibfnamefont {X.-F.}\ \bibnamefont
  {Chen}}, \bibinfo {author} {\bibfnamefont {C.}~\bibnamefont {Greiner}},
  \bibinfo {author} {\bibfnamefont {E.}~\bibnamefont {Wang}}, \bibinfo {author}
  {\bibfnamefont {X.-N.}\ \bibnamefont {Wang}}, \ and\ \bibinfo {author}
  {\bibfnamefont {Z.}~\bibnamefont {Xu}},\ }\href {\doibase
  10.1103/PhysRevC.81.064908} {\bibfield  {journal} {\bibinfo  {journal} {Phys.
  Rev. C}\ }\textbf {\bibinfo {volume} {81}},\ \bibinfo {pages} {064908}
  (\bibinfo {year} {2010})},\ \Eprint {http://arxiv.org/abs/1002.1165}
  {arXiv:1002.1165 [nucl-th]} \BibitemShut {NoStop}%
\bibitem [{\citenamefont {Ma}\ \emph {et~al.}(2019)\citenamefont {Ma},
  \citenamefont {Dai}, \citenamefont {Zhang},\ and\ \citenamefont
  {Wang}}]{Ma:2018swx}%
  \BibitemOpen
  \bibfield  {author} {\bibinfo {author} {\bibfnamefont {G.-Y.}\ \bibnamefont
  {Ma}}, \bibinfo {author} {\bibfnamefont {W.}~\bibnamefont {Dai}}, \bibinfo
  {author} {\bibfnamefont {B.-W.}\ \bibnamefont {Zhang}}, \ and\ \bibinfo
  {author} {\bibfnamefont {E.-K.}\ \bibnamefont {Wang}},\ }\href {\doibase
  10.1140/epjc/s10052-019-7005-z} {\bibfield  {journal} {\bibinfo  {journal}
  {Eur. Phys. J. C}\ }\textbf {\bibinfo {volume} {79}},\ \bibinfo {pages} {518}
  (\bibinfo {year} {2019})},\ \Eprint {http://arxiv.org/abs/1812.02033}
  {arXiv:1812.02033 [nucl-th]} \BibitemShut {NoStop}%
\bibitem [{\citenamefont {Neufeld}(2011)}]{Neufeld:2010xi}%
  \BibitemOpen
  \bibfield  {author} {\bibinfo {author} {\bibfnamefont {R.~B.}\ \bibnamefont
  {Neufeld}},\ }\href {\doibase 10.1103/PhysRevD.83.065012} {\bibfield
  {journal} {\bibinfo  {journal} {Phys. Rev. D}\ }\textbf {\bibinfo {volume}
  {83}},\ \bibinfo {pages} {065012} (\bibinfo {year} {2011})},\ \Eprint
  {http://arxiv.org/abs/1011.4979} {arXiv:1011.4979 [hep-ph]} \BibitemShut
  {NoStop}%
\bibitem [{\citenamefont {Miller}\ \emph {et~al.}(2007)\citenamefont {Miller},
  \citenamefont {Reygers}, \citenamefont {Sanders},\ and\ \citenamefont
  {Steinberg}}]{Miller:2007ri}%
  \BibitemOpen
  \bibfield  {author} {\bibinfo {author} {\bibfnamefont {M.~L.}\ \bibnamefont
  {Miller}}, \bibinfo {author} {\bibfnamefont {K.}~\bibnamefont {Reygers}},
  \bibinfo {author} {\bibfnamefont {S.~J.}\ \bibnamefont {Sanders}}, \ and\
  \bibinfo {author} {\bibfnamefont {P.}~\bibnamefont {Steinberg}},\ }\href
  {\doibase 10.1146/annurev.nucl.57.090506.123020} {\bibfield  {journal}
  {\bibinfo  {journal} {Ann. Rev. Nucl. Part. Sci.}\ }\textbf {\bibinfo
  {volume} {57}},\ \bibinfo {pages} {205} (\bibinfo {year} {2007})},\ \Eprint
  {http://arxiv.org/abs/nucl-ex/0701025} {arXiv:nucl-ex/0701025} \BibitemShut
  {NoStop}%
\bibitem [{\citenamefont {Pang}\ \emph {et~al.}(2016)\citenamefont {Pang},
  \citenamefont {Petersen}, \citenamefont {Wang},\ and\ \citenamefont
  {Wang}}]{Pang:2016igs}%
  \BibitemOpen
  \bibfield  {author} {\bibinfo {author} {\bibfnamefont {L.-G.}\ \bibnamefont
  {Pang}}, \bibinfo {author} {\bibfnamefont {H.}~\bibnamefont {Petersen}},
  \bibinfo {author} {\bibfnamefont {Q.}~\bibnamefont {Wang}}, \ and\ \bibinfo
  {author} {\bibfnamefont {X.-N.}\ \bibnamefont {Wang}},\ }\href {\doibase
  10.1103/PhysRevLett.117.192301} {\bibfield  {journal} {\bibinfo  {journal}
  {Phys. Rev. Lett.}\ }\textbf {\bibinfo {volume} {117}},\ \bibinfo {pages}
  {192301} (\bibinfo {year} {2016})},\ \Eprint
  {http://arxiv.org/abs/1605.04024} {arXiv:1605.04024 [hep-ph]} \BibitemShut
  {NoStop}%
\bibitem [{\citenamefont {Putschke}\ \emph {et~al.}(2019)\citenamefont
  {Putschke} \emph {et~al.}}]{Putschke:2019yrg}%
  \BibitemOpen
  \bibfield  {author} {\bibinfo {author} {\bibfnamefont {J.~H.}\ \bibnamefont
  {Putschke}} \emph {et~al.},\ }\href@noop {} {\  (\bibinfo {year} {2019})},\
  \Eprint {http://arxiv.org/abs/1903.07706} {arXiv:1903.07706 [nucl-th]}
  \BibitemShut {NoStop}%
\bibitem [{\citenamefont {Andersson}\ \emph {et~al.}(1983)\citenamefont
  {Andersson}, \citenamefont {Gustafson},\ and\ \citenamefont
  {Soderberg}}]{Andersson:1983jt}%
  \BibitemOpen
  \bibfield  {author} {\bibinfo {author} {\bibfnamefont {B.}~\bibnamefont
  {Andersson}}, \bibinfo {author} {\bibfnamefont {G.}~\bibnamefont
  {Gustafson}}, \ and\ \bibinfo {author} {\bibfnamefont {B.}~\bibnamefont
  {Soderberg}},\ }\href {\doibase 10.1007/BF01407824} {\bibfield  {journal}
  {\bibinfo  {journal} {Z. Phys. C}\ }\textbf {\bibinfo {volume} {20}},\
  \bibinfo {pages} {317} (\bibinfo {year} {1983})}\BibitemShut {NoStop}%
\bibitem [{\citenamefont {Sjostrand}(1984)}]{Sjostrand:1984ic}%
  \BibitemOpen
  \bibfield  {author} {\bibinfo {author} {\bibfnamefont {T.}~\bibnamefont
  {Sjostrand}},\ }\href {\doibase 10.1016/0550-3213(84)90607-2} {\bibfield
  {journal} {\bibinfo  {journal} {Nucl. Phys. B}\ }\textbf {\bibinfo {volume}
  {248}},\ \bibinfo {pages} {469} (\bibinfo {year} {1984})}\BibitemShut
  {NoStop}%
\bibitem [{\citenamefont {Cooper}\ and\ \citenamefont
  {Frye}(1974)}]{Cooper:1974mv}%
  \BibitemOpen
  \bibfield  {author} {\bibinfo {author} {\bibfnamefont {F.}~\bibnamefont
  {Cooper}}\ and\ \bibinfo {author} {\bibfnamefont {G.}~\bibnamefont {Frye}},\
  }\href {\doibase 10.1103/PhysRevD.10.186} {\bibfield  {journal} {\bibinfo
  {journal} {Phys. Rev. D}\ }\textbf {\bibinfo {volume} {10}},\ \bibinfo
  {pages} {186} (\bibinfo {year} {1974})}\BibitemShut {NoStop}%
\bibitem [{\citenamefont {Casalderrey-Solana}\ \emph
  {et~al.}(2017)\citenamefont {Casalderrey-Solana}, \citenamefont {Gulhan},
  \citenamefont {Milhano}, \citenamefont {Pablos},\ and\ \citenamefont
  {Rajagopal}}]{Casalderrey-Solana:2016jvj}%
  \BibitemOpen
  \bibfield  {author} {\bibinfo {author} {\bibfnamefont {J.}~\bibnamefont
  {Casalderrey-Solana}}, \bibinfo {author} {\bibfnamefont {D.}~\bibnamefont
  {Gulhan}}, \bibinfo {author} {\bibfnamefont {G.}~\bibnamefont {Milhano}},
  \bibinfo {author} {\bibfnamefont {D.}~\bibnamefont {Pablos}}, \ and\ \bibinfo
  {author} {\bibfnamefont {K.}~\bibnamefont {Rajagopal}},\ }\href {\doibase
  10.1007/JHEP03(2017)135} {\bibfield  {journal} {\bibinfo  {journal} {JHEP}\
  }\textbf {\bibinfo {volume} {03}},\ \bibinfo {pages} {135} (\bibinfo {year}
  {2017})},\ \Eprint {http://arxiv.org/abs/1609.05842} {arXiv:1609.05842
  [hep-ph]} \BibitemShut {NoStop}%
\bibitem [{\citenamefont {Acharya}\ \emph {et~al.}(2022)\citenamefont {Acharya}
  \emph {et~al.}}]{ALargeIonColliderExperiment:2021mqf}%
  \BibitemOpen
  \bibfield  {author} {\bibinfo {author} {\bibfnamefont {S.}~\bibnamefont
  {Acharya}} \emph {et~al.} (\bibinfo {collaboration} {A Large Ion Collider
  Experiment, ALICE}),\ }\href {\doibase 10.1103/PhysRevLett.128.102001}
  {\bibfield  {journal} {\bibinfo  {journal} {Phys. Rev. Lett.}\ }\textbf
  {\bibinfo {volume} {128}},\ \bibinfo {pages} {102001} (\bibinfo {year}
  {2022})},\ \Eprint {http://arxiv.org/abs/2107.12984} {arXiv:2107.12984
  [nucl-ex]} \BibitemShut {NoStop}%
\bibitem [{\citenamefont {Aad}\ \emph {et~al.}(2023{\natexlab{a}})\citenamefont
  {Aad} \emph {et~al.}}]{ATLAS:2022vii}%
  \BibitemOpen
  \bibfield  {author} {\bibinfo {author} {\bibfnamefont {G.}~\bibnamefont
  {Aad}} \emph {et~al.} (\bibinfo {collaboration} {ATLAS}),\ }\href {\doibase
  10.1103/PhysRevC.107.054909} {\bibfield  {journal} {\bibinfo  {journal}
  {Phys. Rev. C}\ }\textbf {\bibinfo {volume} {107}},\ \bibinfo {pages}
  {054909} (\bibinfo {year} {2023}{\natexlab{a}})},\ \Eprint
  {http://arxiv.org/abs/2211.11470} {arXiv:2211.11470 [nucl-ex]} \BibitemShut
  {NoStop}%
\bibitem [{\citenamefont {Ehlers}(2022)}]{Ehlers:2022dfp}%
  \BibitemOpen
  \bibfield  {author} {\bibinfo {author} {\bibfnamefont {R.}~\bibnamefont
  {Ehlers}} (\bibinfo {collaboration} {ALICE}),\ }\href {\doibase
  10.22323/1.414.0460} {\bibfield  {journal} {\bibinfo  {journal} {PoS}\
  }\textbf {\bibinfo {volume} {ICHEP2022}},\ \bibinfo {pages} {460} (\bibinfo
  {year} {2022})},\ \Eprint {http://arxiv.org/abs/2211.11800} {arXiv:2211.11800
  [nucl-ex]} \BibitemShut {NoStop}%
\bibitem [{\citenamefont {Aad}\ \emph {et~al.}(2023{\natexlab{b}})\citenamefont
  {Aad} \emph {et~al.}}]{ATLAS:2023hso}%
  \BibitemOpen
  \bibfield  {author} {\bibinfo {author} {\bibfnamefont {G.}~\bibnamefont
  {Aad}} \emph {et~al.} (\bibinfo {collaboration} {ATLAS}),\ }\href {\doibase
  10.1103/PhysRevLett.131.172301} {\bibfield  {journal} {\bibinfo  {journal}
  {Phys. Rev. Lett.}\ }\textbf {\bibinfo {volume} {131}},\ \bibinfo {pages}
  {172301} (\bibinfo {year} {2023}{\natexlab{b}})},\ \Eprint
  {http://arxiv.org/abs/2301.05606} {arXiv:2301.05606 [nucl-ex]} \BibitemShut
  {NoStop}%
\bibitem [{\citenamefont {Kang}\ \emph
  {et~al.}(2025{\natexlab{b}})\citenamefont {Kang}, \citenamefont {Wang},
  \citenamefont {Wang},\ and\ \citenamefont {Zhang}}]{Kang:2023ycg}%
  \BibitemOpen
  \bibfield  {author} {\bibinfo {author} {\bibfnamefont {J.-W.}\ \bibnamefont
  {Kang}}, \bibinfo {author} {\bibfnamefont {S.}~\bibnamefont {Wang}}, \bibinfo
  {author} {\bibfnamefont {L.}~\bibnamefont {Wang}}, \ and\ \bibinfo {author}
  {\bibfnamefont {B.-W.}\ \bibnamefont {Zhang}},\ }\href {\doibase
  10.1103/PhysRevC.111.054905} {\bibfield  {journal} {\bibinfo  {journal}
  {Phys. Rev. C}\ }\textbf {\bibinfo {volume} {111}},\ \bibinfo {pages}
  {054905} (\bibinfo {year} {2025}{\natexlab{b}})},\ \Eprint
  {http://arxiv.org/abs/2312.15518} {arXiv:2312.15518 [hep-ph]} \BibitemShut
  {NoStop}%
\bibitem [{\citenamefont {Aad}\ \emph {et~al.}(2020)\citenamefont {Aad} \emph
  {et~al.}}]{ATLAS:2019iaa}%
  \BibitemOpen
  \bibfield  {author} {\bibinfo {author} {\bibfnamefont {G.}~\bibnamefont
  {Aad}} \emph {et~al.} (\bibinfo {collaboration} {ATLAS}),\ }\href {\doibase
  10.1007/JHEP03(2020)179} {\bibfield  {journal} {\bibinfo  {journal} {JHEP}\
  }\textbf {\bibinfo {volume} {03}},\ \bibinfo {pages} {179} (\bibinfo {year}
  {2020})},\ \Eprint {http://arxiv.org/abs/1912.09866} {arXiv:1912.09866
  [hep-ex]} \BibitemShut {NoStop}%
\end{thebibliography}%

\end{document}